\documentclass[a4paper,fleqn,usenatbib]{mnras}

\usepackage[T1]{fontenc}
\usepackage{ae,aecompl}

\usepackage{graphicx}	
\usepackage{amsmath}	
\usepackage{amssymb}	
\usepackage[table]{xcolor} 
\usepackage{rotating} 

\graphicspath{{./figures_arxiv/}}

\newcommand{\ie}{{\it i.e.}}

\newcommand{\eg}{{\it e.g.}}

\newcommand{\eq}{Eq.}

\newcommand{\fig}{Fig.}
\newcommand{\Fig}{Fig.}
\newcommand{\figs}{Figures}

\newcommand{\Sec}{Section}

\newcommand{\Tab}{Tab.}

\newcommand{\figu}[1]{\fig~\ref{fig:#1}}
\newcommand{\tabl}[1]{\Tab~\ref{tab:#1}}

\newcommand{\first}{SR-$0$S}
\newcommand{\second}{SR-LS}
\newcommand{\third}{WR-MS}
\newcommand{\fourth}{WR-HS}

\newcommand{\revised}[1]{#1}

\hypersetup{draft}

\title[UHECR nuclei in GRB multi-collision models]{Systematic parameter space study for the UHECR origin from GRBs in models with multiple internal shocks}

\author[J. Heinze et al.]{
J. Heinze,$^{1}$\thanks{E-mail: jonas.heinze@desy.de}
D. Biehl,$^{1}$
A. Fedynitch,$^{1,4}$
D. Boncioli,$^{1,2,3}$ 
A. Rudolph$^{1}$
and W. Winter$^{1}$
\\
$^{1}$Deutsches Elektronen-Synchrotron (DESY), Platanenallee 6, D-15738 Zeuthen, Germany\\
$^{2}$Universit\`{a} dell'Aquila, Dipartimento di Scienze Fisiche e Chimiche, L'Aquila, Italy\\
$^{3}$INFN Laboratori Nazionali del Gran Sasso, Assergi (L'Aquila), Italy\\
$^{4}$Institute  for  Cosmic  Ray  Research,  the  University  of  Tokyo, 5-1-5  Kashiwa-no-ha,  Kashiwa,  Chiba  277-8582,  Japan
}

\date{Accepted XXX. Received YYY; in original form ZZZ}

\pubyear{2020}

\begin{document}
\label{firstpage}
\pagerange{\pageref{firstpage}--\pageref{lastpage}}
\maketitle

\begin{abstract}
We scrutinize the paradigm that conventional long-duration Gamma-Ray Bursts (GRBs) are the dominant source of the ultra-high energy cosmic rays (UHECRs) within the internal shock scenario by describing UHECR spectrum and composition and by studying the predicted (source and cosmogenic) neutrino fluxes. Since it has been demonstrated that the stacking searches for astrophysical GRB neutrinos strongly constrain the parameter space  in single-zone models, we focus on the  dynamics of multiple collisions for which different messengers are expected to come from different regions of the same object. We propose a model which can describe both stochastic and deterministic engines, which we study in a systematic way. We find that GRBs can indeed describe the UHECRs for a wide range of different model assumptions with comparable quality \revised{albeit with the previously known problematic energy requirements}; the heavy mass fraction at injection is found to be larger than 70\% ($95 \%$ CL).
We demonstrate that the post-dicted (from UHECR data) neutrino fluxes from sources and UHECR propagation are indeed below the current sensitivities but will be reached by the next generation of experiments. We finally critically review the required source energetics with the specific examples found in this study.
\end{abstract}




\section{Introduction}

Gamma-Ray Bursts (GRBs) have been proposed to be powerful enough to describe Ultra-High Energy Cosmic Rays (UHECRs) \revised{\citep{waxman1995cosmological, vietri1995acceleration}}. The interactions of cosmic rays with photons in the source can lead to substantial neutrino production during the prompt emission phase of gamma rays, see \eg\ \citet{Waxman:1997ti, Murase:2005hy,Hummer:2011ms}. In this work, we focus on conventional long-duration GRBs with isotropic luminosities around $10^{52.5} \mathrm{erg \, s^{-1}}$~\citep{Gruber:2014iza}. The searches for GRB neutrinos are highly sensitive because the directional, timing and energy information can be used to suppress background. Therefore, the neutrino flux limits from the IceCube Observatory for GRBs are the best among all potential source classes~\citep{Abbasi:2012zw,Aartsen:2017wea}. Besides that, the observation of a temporally and spatially coincident neutrino emission is one of the few ways to locate transient cosmic ray accelerators such as GRBs. So far, no source neutrinos from GRBs have been observed. The interaction of cosmic rays with the cosmic backgrounds during propagation to Earth can also lead to a ``cosmogenic'' neutrino flux \citep{Beresinsky:1969qj,Aab:2016zth,AlvesBatista:2018zui,Heinze:2019jou} but no cosmogenic neutrinos have been found either \citep{Aartsen:2018vtx,Aab:2019auo}. 
A model-dependent correlation analysis of the UHECR arrival directions with source catalogs, performed by the Pierre Auger Observatory for Active Galactic Nuclei and Starburst Galaxies, indicates an intermediate-scale anisotropy related to the distribution of nearby UHECR sources \citep{Aab:2018chp}. According to this result, only a limited fraction of the observed UHECR flux can be attributed to known starburst galaxies or active galactic nuclei, leaving the main contribution to the UHECR flux unconstrained. GRBs may be therefore a conceivable option for the remaining part.

In the present study, we follow the hypothesis that GRBs are the dominant sources of the UHECRs and describe the UHECR spectrum and composition within the internal shock scenario \revised{\citep{Rees:1994nw, Kobayashi:1997jk, Daigne:1998xc}}. It has been shown that simple one-zone-emission models are in tension with the neutrino flux limits for most of the parameter space, both for the case of a pure proton composition~\citep{Baerwald:2014zga} and for a heavier primary composition~\citep{Biehl:2017zlw}. This argument follows from the over-production of source neutrinos due to high radiation densities in the source~\citep{Biehl:2017zlw} for luminosities and collision radii typically assumed for GRBs -- which points towards alternative sources, such as low-luminosity GRBs~(\citet{Zhang:2017moz,Boncioli:2018lrv}, see also \citet{Murase:2008mr, Samuelsson:2018fan}).

Multi-collision models, such as the one discussed in this work, exhibit a lower neutrino flux~\citep{Bustamante:2014oka,Globus:2014fka}, which is typically dominated by cosmic-ray interactions close to the photosphere where the radiation densities are high. If only a small fraction of the collisions occurs there, the overall neutrino flux is lower than in the single zone model for the same proton injection luminosity. For a discussion of the production regions of cosmic messengers, the impact of the engine properties on the different messengers and light curves in different energy bands, and the dependence on the collision model for the shells, see \citet{Bustamante:2014oka,Bustamante:2016wpu,Rudolph:2019ccl}.

In contrast to one-zone models, multi-collision models add the possibility of studying different jet structures and corresponding light curves. A multi-peaked light curve can be generated by a smooth Lorentz factor variation in the outflow, short-time variability by additional random variations (\textit{stochasticity}) in the Lorentz factor profile. A discussion on different outflow structures and corresponding light curves can be found in \eg \, \citet{Daigne:1998xc, Bosnjak:2008bd}). Also \citet{Bustamante:2016wpu} investigate different Lorentz factor profiles in connection with neutrino and gamma-ray light curves for proton-loaded jets, however without fitting the source parameters to data. Most GRB multi-collision models do not jointly describe the UHECR spectrum and composition -- except \citet{Globus:2014fka,Globus:2015xga}, who performed a complete investigation of the multi-collision model in the context of UHECR including nuclei for a smooth, continuous outflow that corresponds to a single-peaked light curve without a short-time variability or an intermittent engine. They draw a self-consistent picture for one set of parameters and one collision model, with a neutrino flux prediction close to the current stackin    g limit. No substantial progress on the subject has been made since then, owing to the computational complexity of the problem.

In \Sec~\ref{sec:GRB_model_section} of this paper, we present our multi-collision GRB model for different outflow patterns accounting for a smooth temporal profile (a ramp-up of the Lorentz factor towards later times) in combination with a stochastic engine behavior, mainly based on an extension of the techniques developed in our previous works. \Sec~\ref{sec:fitting_to_data} describes the systematic scan over the engine parameters and how these are estimated from UHECR data. For the first time, we obtain contours on the allowed parameter space from UHECR data and discuss the result in detail based on four representative cases. \Sec~\ref{sec:multi-messenger-results} is dedicated to the multi-messenger signatures, \ie{} the light-curves and the neutrino flux predictions. We aim to conclude whether the absence of associations of IceCube neutrinos with GRBs \cite{Abbasi:2012zw} really disfavors, or even excludes, those as a dominant source of UHECRs when a more sophisticated model for the multi-messenger production is used.

\section{GRB source model}
\label{sec:GRB_model_section}

\subsection{Multi-collision dynamics}
\label{sec:GRB_model}

 We follow the model first presented in \citet{Kobayashi:1997jk} using the implementation described on detail in \citet{Bustamante:2016wpu, Rudolph:2019ccl} in order to describe the relativistic outflow of the GRB jet. We briefly review the model here  and discuss a  few modifications and improvements to the model implemented since then.
We also discuss the key differences to the collision model from \citet{Globus:2014fka}, which is based on \citet{Daigne:1998xc}.

The relativistic outflow with non-uniform density and velocity profiles is approximated by a series of plasma shells with \revised{distinct masses $m_k$, Lorentz factors $\Gamma_k$ and widths $l_k$, separated by distances $d_k$}. Due to the gradient in velocity, the more rapid shells will eventually catch up to slower shells. A collision between a rapid shell (index $r$) and a slow shell (index $s$) forms a new, merged shell (index $m$).
Due to energy conservation, the energy dissipated during the collision is given by the difference of kinetic energy before and after the collision:
\begin{align}
 E_{\rm C} = m_r \Gamma_r + m_s \Gamma_s - (m_r + m_s) \Gamma_m \, .
\end{align}
Momentum and mass conservation implies that the Lorentz factor of the merged shell $\Gamma_m$ is given by: 
\begin{align}
	\label{eq:gamma_merged}
 \Gamma_m \simeq \sqrt{\frac{m_r \Gamma_r + m_s \Gamma_s}{m_r/ \Gamma_r + m_s/ \Gamma_s}} \, . 
\end{align}
In \citet{Daigne:1998xc}, most of the energy is dissipated during the initial phase where the less massive shell sweeps up a mass equal to its own in the other shell, such that $\Gamma_\mathrm{C} = \sqrt{\Gamma_r \cdot \Gamma_s}$. We follow this assumption here, diverting from \citet{Bustamante:2016wpu}.
Individual and merged shells continue to propagate in the fireball until they collide with other shells or reach the circumburst medium, where they are taken out of the simulation. \revised{} The circumburst medium is assumed to dominate the outflow dynamics at a distance $R_\mathrm{circumburst} \gtrsim 5.5 \ \cdot 10^{16} \ \mathrm{cm}$ from the central engine. 
\revised{For simiplicity we do not calculate of the deceleration radius for each fireball explicitly, but instead impose this general value compatible with theoretical estimates \citep{Rees:1992ek} and values inferred from observations \citep{Liang:2009zi}}.

In each collision particles are accelerated and emitted. We assume that the timescale of emission $\delta t_{\rm em}$ is given by the time the reverse shock takes to cross the rapid shell. Since both shells get compressed during the collision process, the width of the merged shell is reduced during the collision ($l_{m, \mathrm{C}} < l_s + l_r$). The a full derivation for the formulas for $\delta t_{\rm em}$ and $l_{m, \mathrm{C}}$ is contained in \citet{Kobayashi:1997jk, Bustamante:2016wpu}. In a different scenario, the internal energy remaining in the shell after the collision will result in an expansion of the merged shell, which is discussed in more detail in \citet{Rudolph:2019ccl}. Here we assume that the merged shell after the collision recovers the width of a single shell before the collision ($l_{m, \mathrm{final}} = l_s = l_r$). At all times, the shells contained in the fireball thus share the same width.
In \citet{Globus:2014fka}, shell widths do not play a role since the energy densities are computed assuming a constant kinetic wind luminosity. This approach is strictly valid only for the first round of collisions and roughly corresponds to the assumption of a constant shell width.

The final emitted particle spectra are computed by summing over all single collision spectra. A collision that occurs at time $t_{\rm C}$ at a distance $R_{\rm C}$ from the engine will start to be observed at time 
 $t_{\rm obs} = t_{\rm C} - R_{\rm C}/c$. 

We calculate the time-dependent flux from each collision assuming a 'Fast Rise and Exponential Decay' (FRED) shape normalized to the total gamma-ray luminosity of the collision (see \citet{Kobayashi:1997jk, Bustamante:2016wpu} for details).

\paragraph*{Initial setup}

The simulations start with 1000 shells with equal kinetic energies $E_{\rm kin} = 10^{52}~\mathrm{erg}$,\footnote{If not noted otherwise, energies are isotropic-equivalent energies.} shell widths $l = c \cdot 0.002~\mathrm{s}$ and shell separations $d = c \cdot 0.002~\mathrm{s}$. The innermost shell is assumed to be located at radius $R_{\rm min} = 10^8~\mathrm{cm}$\revised{, which is much smaller than the $10^{13}$~cm where the first collisions occur.} After the fireball simulation completes, the kinetic energy $E_{\rm kin}$ is re-normalized to ensure a gamma-ray output $E_\gamma \simeq 10^{53} \, \mathrm{erg}$ over all collisions.

The initial shell distribution follows a log-normal distribution around a deterministic (temporal) Lorentz factor profile $\Gamma_{0,k}$
that assigns higher velocities to shells emitted at later stages of the fireball evolution: \begin{align}
\Gamma_{0,k} = \begin{cases}
		\frac{\Gamma_\mathrm{max} + \Gamma_\mathrm{min}}{2} - \frac{\Gamma_\mathrm{max} - \Gamma_\mathrm{min}}{2} \cdot \cos(\pi \cdot \frac{k}{0.4 \cdot N_\mathrm{shells}})\\
		\qquad \mathrm{if} \quad k <= 0.4 \cdot N_\mathrm{shells}\\
		\Gamma_\mathrm{max} \\
		\qquad \mathrm{if} \quad k > 0.4 \cdot N_\mathrm{shells}
	\end{cases} \, .
\end{align} 
\revised{This choice is known to reproduce the smooth and broad peak structure observed in many GRB light curves. Although a deceleration at the end is not forbidden in this model, it would not contribute to the emission since slow shells cannot catch up and interact with the preceding outflow.}
On top of that, the Lorentz factor of the $k$-th shell is assumed to be stochastically distributed around that profile by
\begin{align}
	\label{eq:lorentz_sampled_stochastic}
	\ln \left( \frac{\Gamma_k - 1}{\Gamma_{0,k - 1}}\right) = A_\Gamma \,\cdot x \, ,
\end{align}
where $x$ is sampled from a Gaussian $P(x) dx = \exp(-x^2)/\sqrt{2 \pi} dx$. The spread $A_\Gamma$ describes the strength of the fluctuations, and 
the values of $\Gamma_\mathrm{min}$ and $\Gamma_\mathrm{max}$ control the mean and dynamic range of the entire distribution. \revised{Since shells are assumed to have equal energy in our model,} the bulk Lorentz factor is calculated as
\begin{align}
	\Gamma_\mathrm{bulk} 
	= \langle E_\mathrm{kin} \rangle / \langle M \rangle
	= N_\mathrm{shells} \left[\sum_k 1/\Gamma_k \right]^{-1} .
\end{align}
This depends non-linearly on the parameters $\Gamma_\mathrm{min}$, $\Gamma_\mathrm{max}$ and $A_\Gamma$. \revised{In order to reduce the number of free parameters the initial shell widths and separations are assumed to be equal for all shells ($d_k = l_k = c \cdot 0.002~\mathrm{s}$). The shell masses are then defined as $m_k = E_{\rm kin} / \Gamma_k$ due to the equal kinetic energy $E_{\rm kin}$ per shell.}

\paragraph*{Benchmark engine profiles}
\label{par:examples_description}

\begin{figure*}
	\begin{tabular}{cccc}
		\textbf{\first}&
		\textbf{\second}&
		\textbf{\third}&
		\textbf{\fourth}\\
		Strong (engine) ramp-up, & 
		Strong (engine) ramp-up,  &
		Weak (engine) ramp-up, &
		Weak (engine) ramp-up, \\
		no stochasticity & 
		low stochasticity &
		medium stochasticity &
		high stochasticity \\
		\includegraphics[width=.22\textwidth]{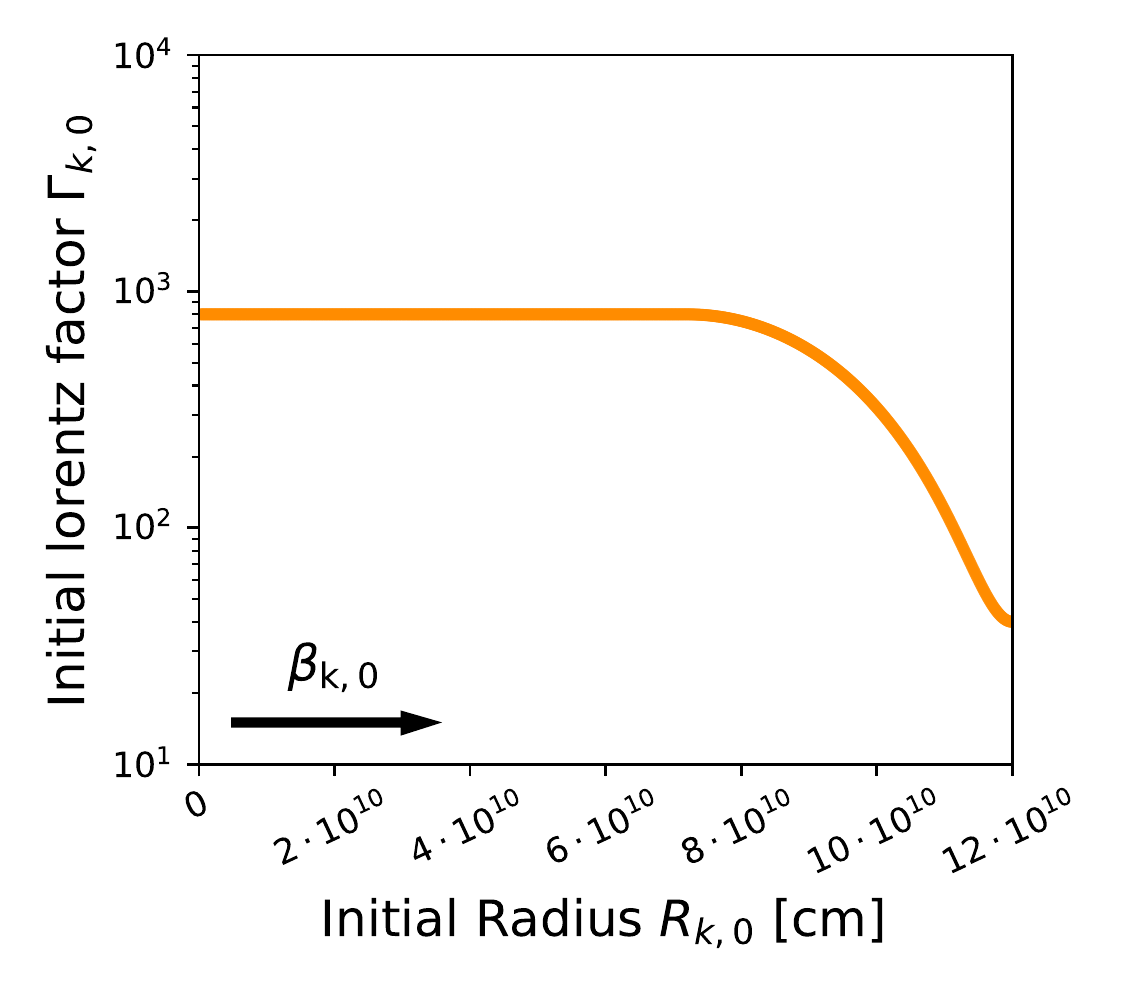} &
		\includegraphics[width=.22\textwidth]{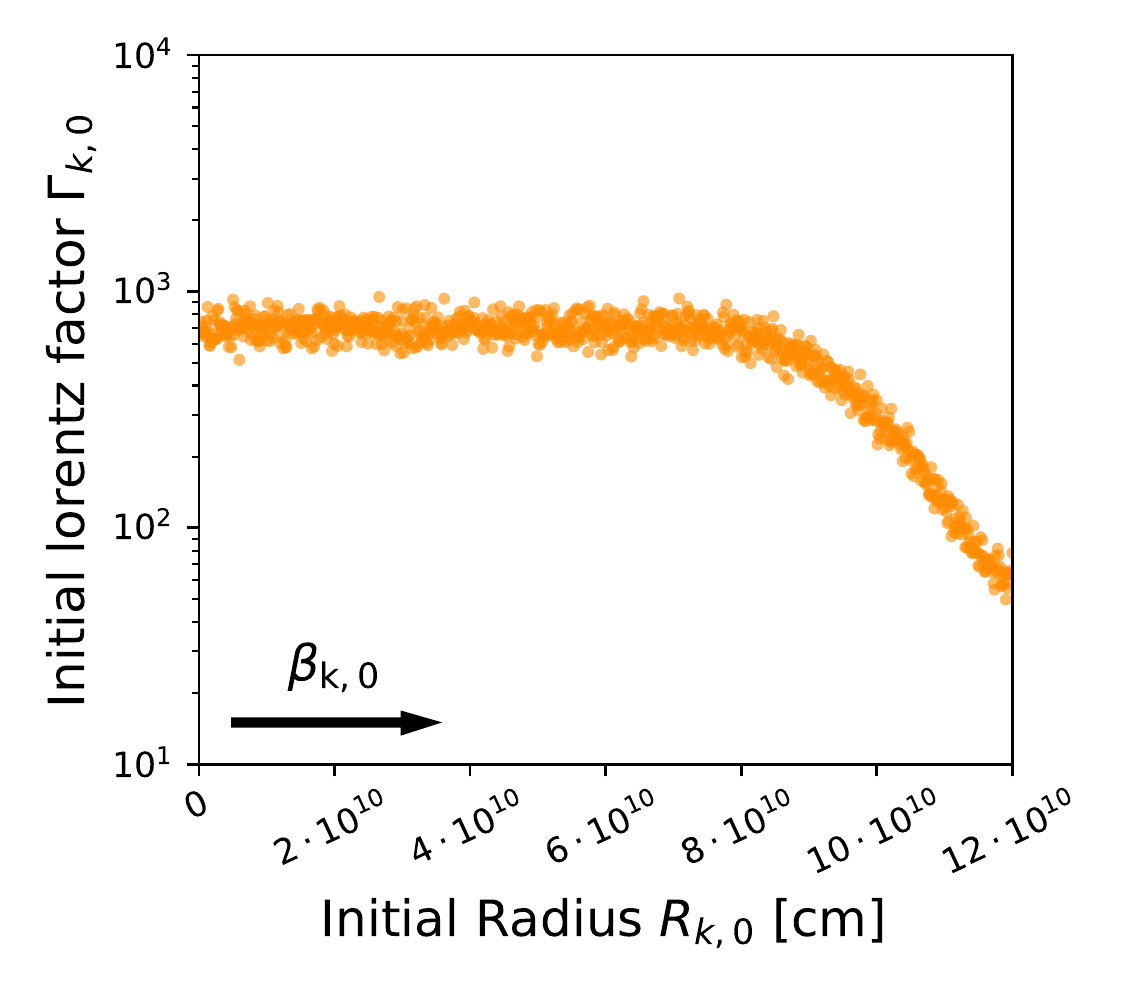} &
		\includegraphics[width=.22\textwidth]{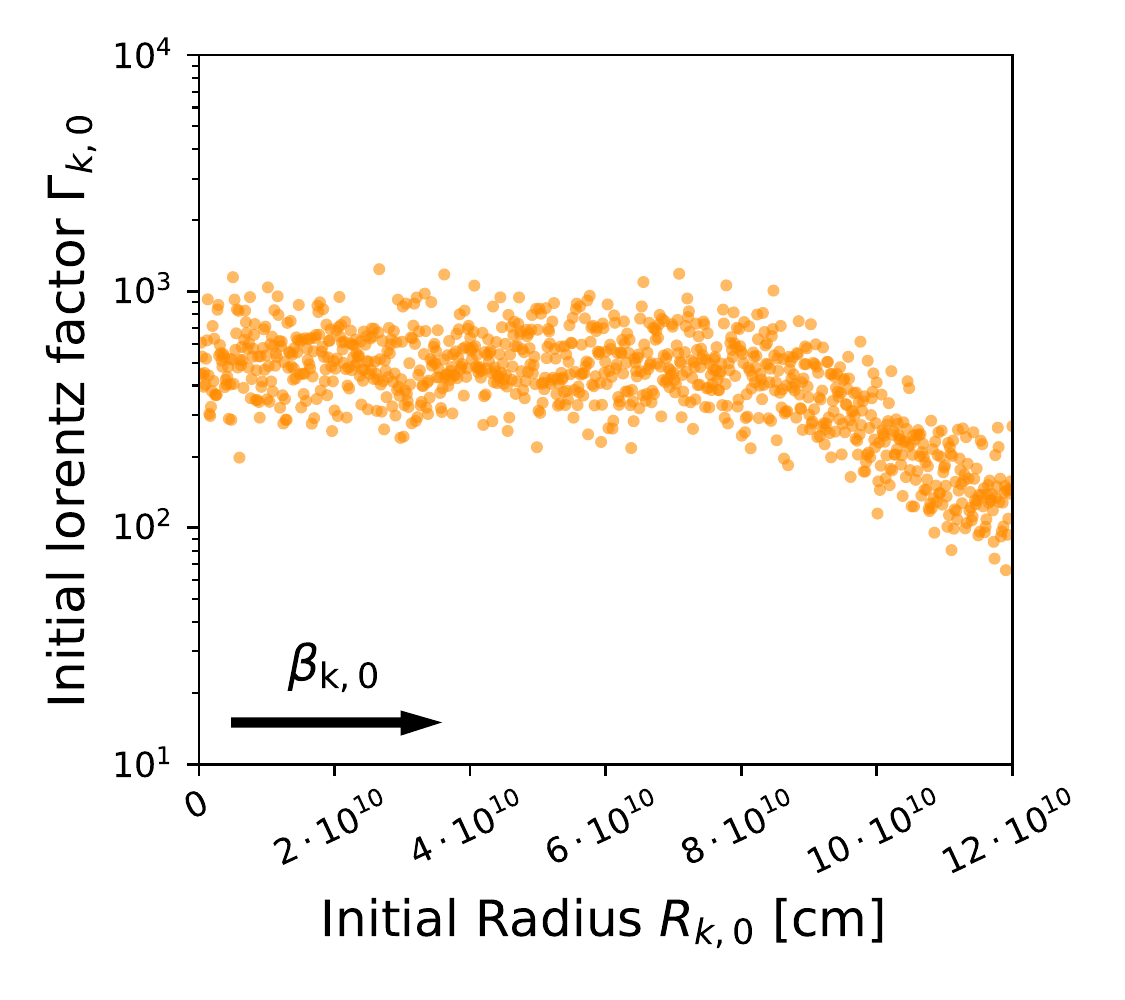} &
		\includegraphics[width=.22\textwidth]{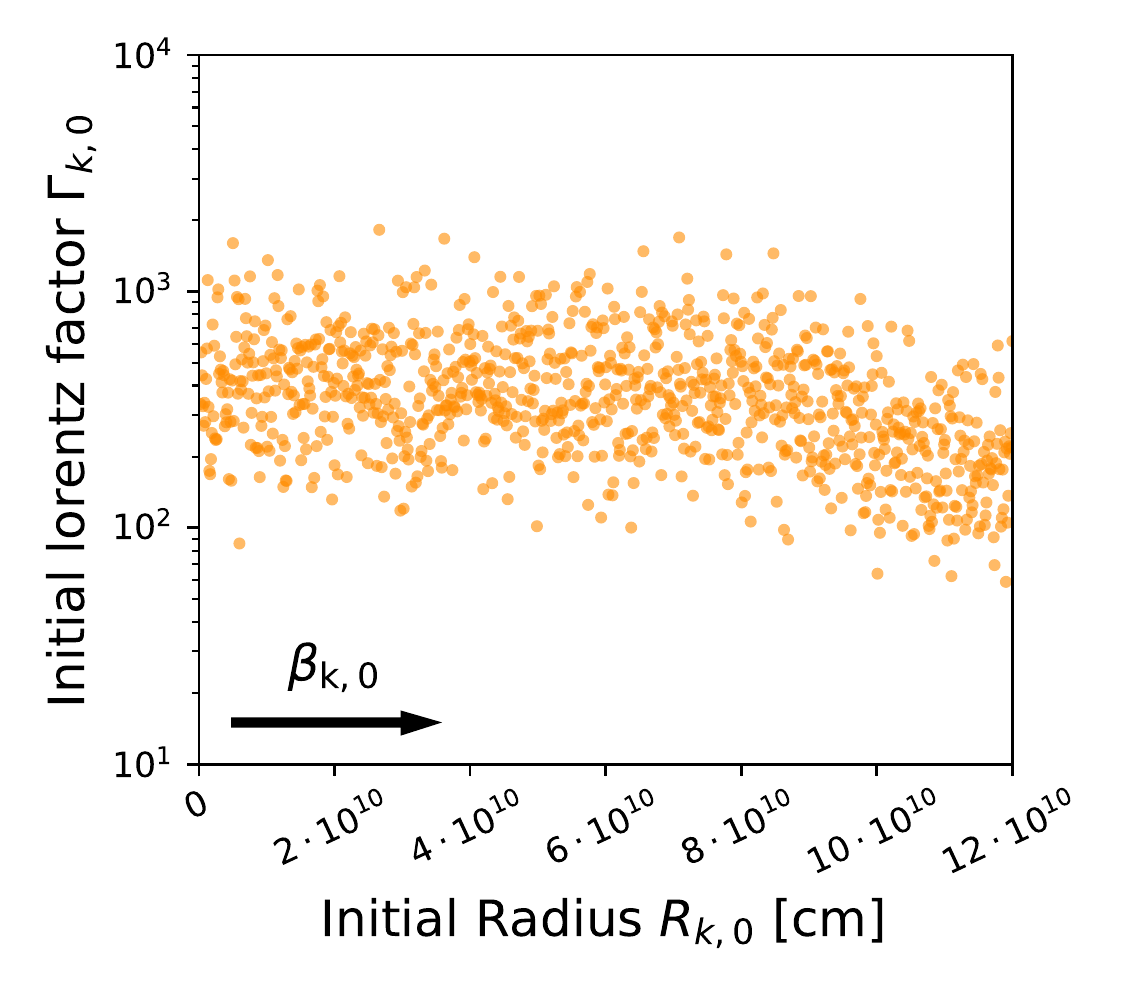}\\
		$\Gamma_\text{min}$: 40,
		$\Gamma_\text{max}$: 800,
		$A_\Gamma$: $0.0$
		& 
		$\Gamma_\text{min}$: 60, 
		$\Gamma_\text{max}$: 700,
		$A_\Gamma$: $0.1$
		& 
		$\Gamma_\text{min}$: 120,
		$\Gamma_\text{max}$: 500,
		$A_\Gamma$: $0.3$
		& 
		$\Gamma_\text{min}$: 160,
		$\Gamma_\text{max}$: 400,
		$A_\Gamma$: $0.5$\\
	\end{tabular}
\caption{\label{fig:initial_shells}Distribution of initial shells for the four example cases, naming convention, and chosen parameters. }
\end{figure*}
In order to simplify the discussion and the visualization of our results, we define four distinct initial benchmark shell configurations. With the formulae from the last section those correspond to four choices of $\Gamma_{\rm min}$, $\Gamma_{\rm max}$ and $A_\Gamma$ that define the strength of the ramp-up and the stochasticity of the engine.

The stochastic fluctuations of the Lorentz factors  are considered to be related to the short time variability on top of a pulsed light curve and have been studied in the fireball framework in the past, see \eg \ \citet{Daigne:1998xc}. \citet{Globus:2014fka} examine the case of a disciplined engine, adding no stochasticity to the ramp-up structure. In addition to the former works, we study the impact of stochastic engine behavior and the strength of the ramp-up in the multi-messenger context including a fit to the observed UHECRs.

The four cases are shown in \fig~\ref{fig:initial_shells} and, sorted by increasing stochasticity, correspond to:        
\begin{enumerate}
 \item[(1)] \textit{Strong ramp-up, no stochasticity (\first )}: Strong ramp-up of the mean Lorentz factor towards later times (thus smaller radii), no stochasticity.
 \item[(2)] \textit{Strong ramp-up, low stochasticity (\second )}: Strong ramp-up of the mean Lorentz factor, some stochasticity. The shell distribution is mainly dominated by the ramp-up. 
 \item[(3)] \textit{Weak ramp-up, medium stochasticity (\third )}: Moderate ramp-up of the mean Lorentz factor, medium stochasticity. The profile and the stochastic features are both pronounced. 
 \item[(4)] \textit{Weak ramp-up, high stochasticity (\fourth )}: Weak ramp-up of the mean Lorentz factor toward later times, high stochasticity. The shell distribution is dominated by the stochasticity, the structure of the profile $\Gamma_{0,k}$ is barely visible.
\end{enumerate}
The initial Lorentz factors are illustrated in \figu{initial_shells} where we also display the corresponding values of $\Gamma_{\rm min}$, $\Gamma_{\rm max}$ and $A_\Gamma$.

\subsection{Radiation model}
\label{sec:rad_model}
The radiation produced by each collision is computed independently using the time-dependent radiation code NeuCosmA~\citep{Biehl:2017zlw}. The total spectrum is then obtained as the sum over all collisions.
The energy dissipated in a collision is distributed among cosmic rays ($\epsilon_{CR}$), electrons ($\epsilon_e$) and magnetic field ($\epsilon_B$) such that $\epsilon_{\mathrm{CR}} + \epsilon_B + \epsilon_e = 1$ and injected into the radiation model. We define the baryonic loading as $f_b = 1/f_e = \epsilon_{\mathrm{CR}} / \epsilon_e$ and calculate the magnetic field following \citet{Bustamante:2016wpu} assuming equipartition between photons and the magnetic field ($\epsilon_B \sim \epsilon_e$).

\revised{Assuming fast cooling of electrons (which is plausible due to the generally high magnetic fields and necessary to efficiently convert internal energy into radiation), each collision deposits $E_{\gamma, \mathrm{C}}= \epsilon_e E_{\mathrm{C}}$ in gamma rays.}
We do not explicitly model the photon fields but instead assume a fixed shape resembling observations: A broken power law peaking at $\epsilon^\prime = 1 \ \mathrm{keV}$ (primed indices refer to the comoving frame of the emitting material) with spectral indices $\alpha = -1$ and $\beta = -2$. We do not explicitly model the target photon spectrum here, but instead postulate that observations are described by the underlying radiation model. More explicit radiative modeling of the photon fields, such as the one in \citet{Globus:2014fka}, would lead to a dependence of the (synchrotron) peak energy on the collision parameters, namely the magnetic field and dissipated energy per mass. For an engine as \first , we expect a simple evolution from higher to lower peak energies as collisions move outside. Stochasticity in the Lorentz factor distribution is expected to add a (stochastic) spread in the distribution of peak energies. The more detailed treatment of these effects goes beyond the scope of this study and would additionally require a radiation model for the electromagnetic processes in the presence of hadronic interactions.
We ensure that the maximal photon energy is high enough in order not to impact the multi-messenger production unless the optical thickness to pair-production exceeds unity, when we impose a cutoff there.
For detailed modeling of GRB spectral energy distribution in the internal shock model see \eg{} \citet{Bosnjak:2008bd, Daigne:2010fb, Bosnjak:2014hya}.

We simulate the nuclear system with the time-dependent radiation code NeuCosmA that iteratively solves the transport equations in order to calculate the cosmic-ray spectra for each collision, see~\citet{Boncioli:2016lkt,Biehl:2017zlw} for details, 
we assume the {\sc Talys} \citep{Koning:2007} disintegration model.
Photo-nuclear processes populate a large variety of secondary elements that are explicitly included within the solver. 
Motivated by Fermi acceleration, nuclei are injected with  $dN' / dE_{\mathrm{CR}}^\prime \propto (E_{\mathrm{CR}}^\prime)^{-2} \exp \left( -E_{\mathrm{CR}}^\prime / E_{\mathrm{CR}, \mathrm{max}}^\prime \right)$.  
The maximum energy $E_{\mathrm{CR}, \mathrm{max}}^\prime$ for each nucleus is determined as in \citet{Biehl:2017zlw} by balancing acceleration with losses due to photo-hadronic interactions, photo-disintegration, photo-pair production, synchrotron emission and adiabatic expansion of the emitting material. For the acceleration, the maximal efficiency of $\eta = 1$ is assumed (Bohm limit). Instead of the interaction timescale $t_\mathrm{int}$ in \citet{Biehl:2017zlw}, we assume the effective energy loss timescale $t'_\mathrm{loss} = A \cdot t'_\mathrm{int}$ limits the maximal energy for photo-disintegration, which is a rough estimate assuming that a single nucleon is ejected per interaction. This assumption is justified because we only consider different mass groups for the injection, namely  hydrogen, helium, nitrogen, silicon and iron, which means nearby isotopes are not distinguished. Protons are limited by the effective cooling timescale for photo-meson production.
The integral fractions of the injection elements are defined as
\begin{equation}
 I_A \equiv \frac{\int\limits_{1 \, \mathrm{GeV}}^{\infty} \frac{dN'}{dE'_{\mathrm{CR}}} E_{\mathrm{CR}}^\prime dE_{\mathrm{CR}}^\prime}{\sum\limits_A \int\limits_{1 \, \mathrm{GeV}}^{\infty} \frac{dN'}{dE'_{\mathrm{CR}}} E_{\mathrm{CR}}^\prime dE_{\mathrm{CR}}^\prime} \, .
 \label{equ:frac}
\end{equation}
They are free parameters of the simulation and are determined by the fit to UHECR data later. \revised{Thus, the fit result resembles an effective or average composition since the initial mass fractions do not explicitely depend on the initial shell radius.} We define the {\em heavy mass fraction} (HMF) as $(I_{\mathrm{N}}+I_{\mathrm{Si}}+I_{\mathrm{Fe}})/(I_{\mathrm{H}}+I_{\mathrm{He}}+I_{\mathrm{N}}+I_{\mathrm{Si}}+I_{\mathrm{Fe}})$. 

Particle injection is assumed to persist throughout the dynamical timescale $t'_\mathrm{dyn} = c \cdot l_{m,\mathrm{C}}^\prime$. 
The injection luminosity is accordingly normalized to $L'_{\mathrm{inj}} = E'_{\mathrm{C}} / t'_{\mathrm{dyn}} $ and the  radiation densities are computed by distributing that luminosity/energy over an isotropic volume $V' = 4 \pi R_{\mathrm{C}}^2 l_{m,\mathrm{C}}^\prime$. The system is evolved over the dynamical timescale $t'_\mathrm{dyn}$, at the end of which the spectra are extracted. 
Note that this assumption is slightly different from \citet{Daigne:1998xc, Globus:2014fka}, where the corresponding time scale is the expansion timescale $t'_{\rm ex}$.

In order to compute the emitted particle spectra, additional assumptions on the escape mechanisms have to be made. In \citet{Baerwald:2013pu,Biehl:2017zlw} neutral particles free-stream while charged particles escape if the edge of the emitting region is within their Larmor radius. This yields an effective escape rate which is comparable to a Bohm-like diffusion process. However, the simulations presented in \citet{Globus:2014fka} suggest a different behavior more similar to a high-pass filter, leading harder, bell-shape escape spectra. Similarly hard escape spectra were analytically derived in \citet{Ohira:2009rd} as $\propto \exp(- \ln^2 (E / E_\mathrm{max}))$, where the escape is most efficient at the maximal energy $E_\mathrm{max}$. For this study, we employ the analytical form derived in \citet{Ohira:2009rd} that yields similar results to \citet{Globus:2014fka} -- supported by the argument that it well describes UHECR data.

A sizable fraction of collisions may occur below the photosphere defined by $\tau^\prime_{\rm Th} = 1$ (where $\tau^\prime_{\rm Th}$ is the optical depth to Thomson scattering) for each individual shell collision, see App.~A.4 in \citet{Bustamante:2016wpu} for details. 
Sub-photospheric collisions, in principle, may lead to high neutrino fluxes if the observed photon spectrum is simply extrapolated to below the photosphere. 
\revised{However, below the photospere shocks are radiation mediated and cosmic rays are not subjected to an abrupt large bulk velocity difference, since their mean free path is shorter than the shock transition region. In consequence, Fermi acceleration is hindered except for a fine tuned combination of conditions \citep{Sironi:2010rb, Beloborodov:2016jmz}, and the target photon spectrum almost certainly has a different shape. }
A simple extrapolation of the radiation model that we apply for the prompt phase therefore tends to overestimate the expected neutrino emission when assuming similar acceleration efficiency. The absence of neutrino observations from GRBs in IceCube may indicate that if GRBs are indeed sources of UHECR, the contribution of sub-photospheric collisions must not be very large in order not to violate the current limits. 
\revised{In light of the reasons set out above, we omit particle emission from sub-photospheric collisions in our model and only take into account collisions occurring in the optically thin regime.}
To avoid an unphysical bias in the parameter scan due to this penalty, we exclude models with less than 40\% of energy dissipated above the photosphere from the interpretation of the parameter scan.

\subsection{Multi-messenger emission regions}
\label{sec:mm_regions}

\begin{figure*}
	\includegraphics[width=.44\textwidth]{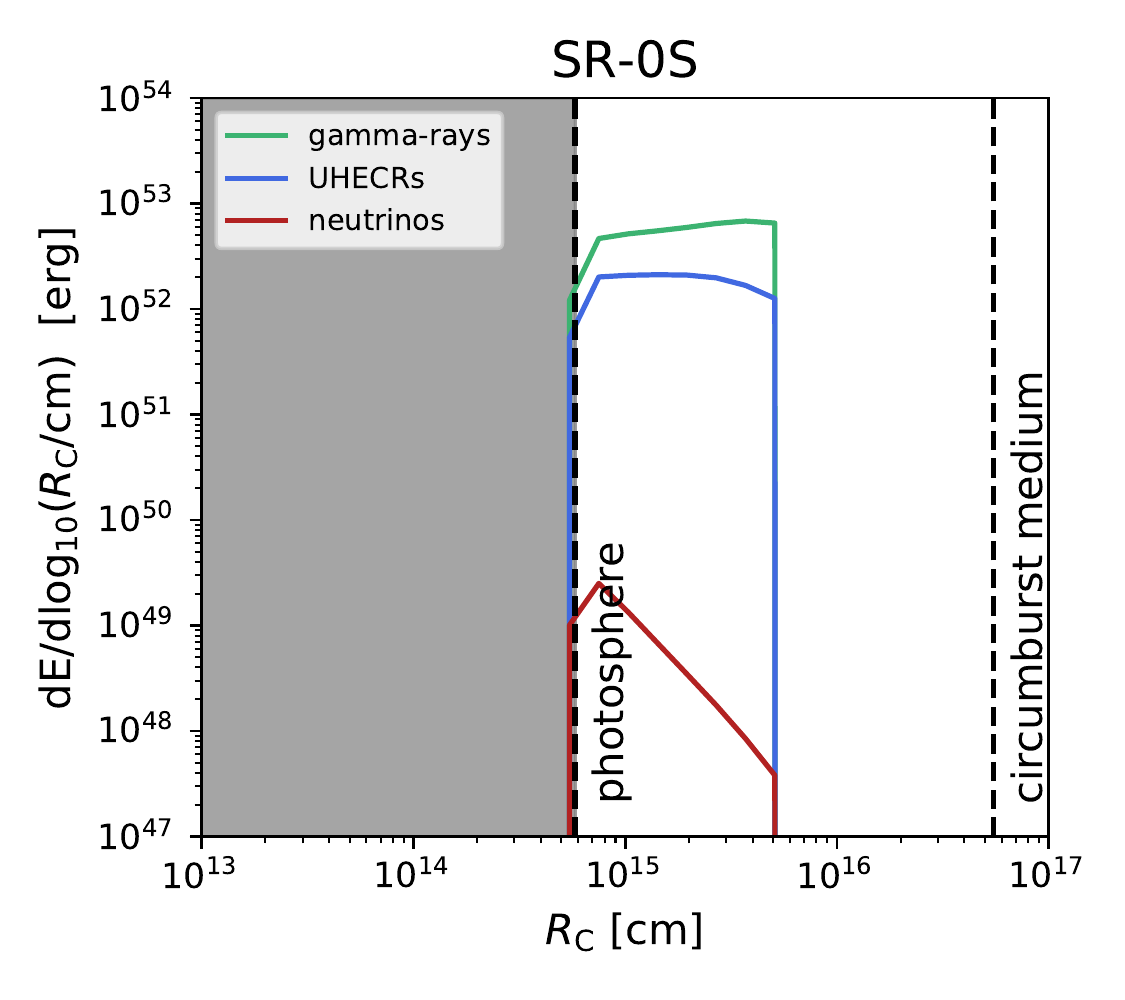}
	\includegraphics[width=.44\textwidth]{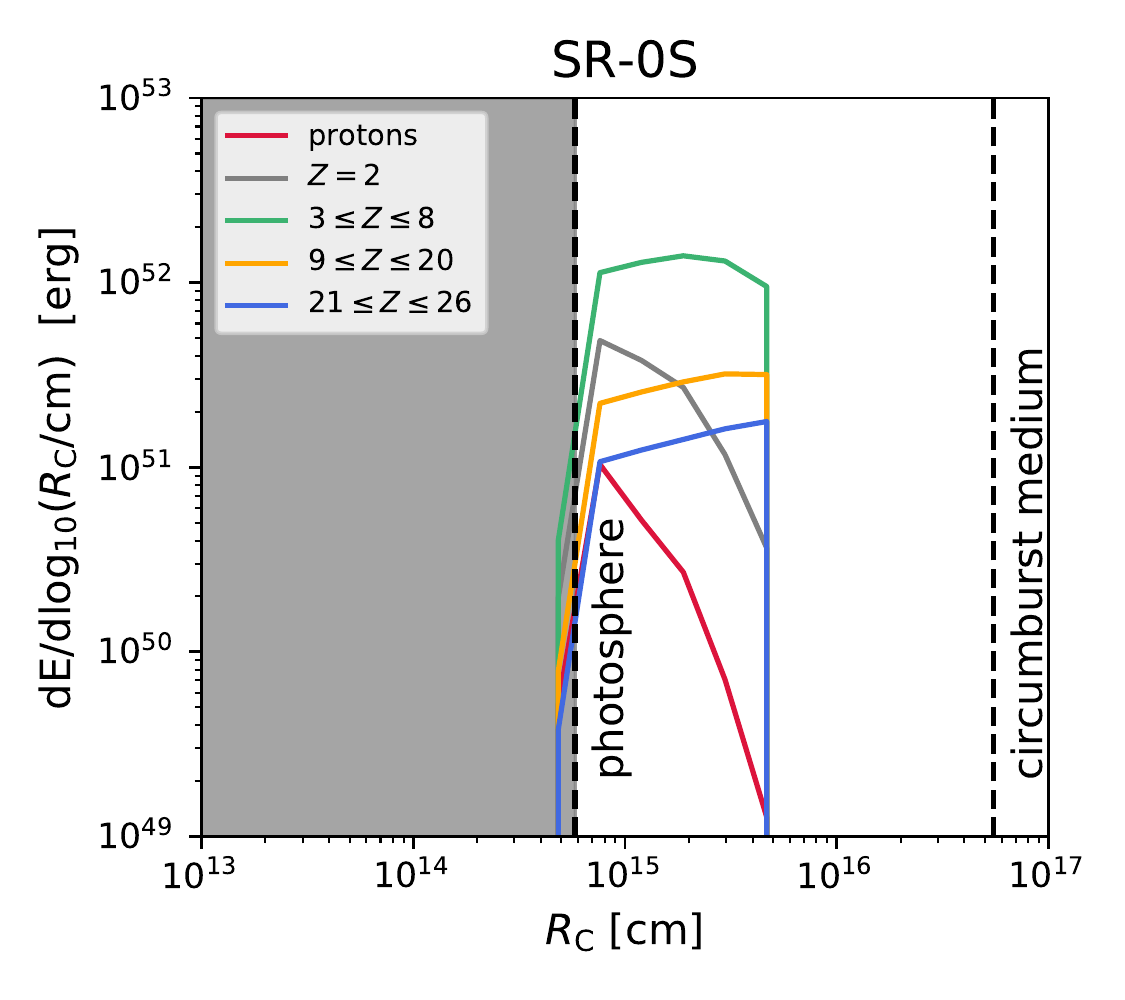}
\caption{\label{fig:particle_radii_first} Energy dissipation in different messengers as a function of $R_{\mathrm{C}}$ for \first{}. The left panel shows different messengers (red: neutrinos, blue: UHECRs, green: gamma rays), the right panel different mass groups. 
In both panels, the dark-shaded areas mark the range in which only subphotospheric collisions occur. The solid curves show the energy dissipated by superphotospheric collisions. The dashed curves illustrate a simple extrapolation of the emission for subphotospheric collisions using the same target photon assumption as beyond the photosphere.  The injected (integral) composition is here fixed to H:~10\% ; He:~25\%, N:~50\%, Si:~10 \%, Fe:~5\%. The output UHECRs are integrated for energies $E_\mathrm{CR} > 10^{10}~\mathrm{GeV}$.}
\end{figure*}

\begin{figure*}
	\includegraphics[width=.44\textwidth]{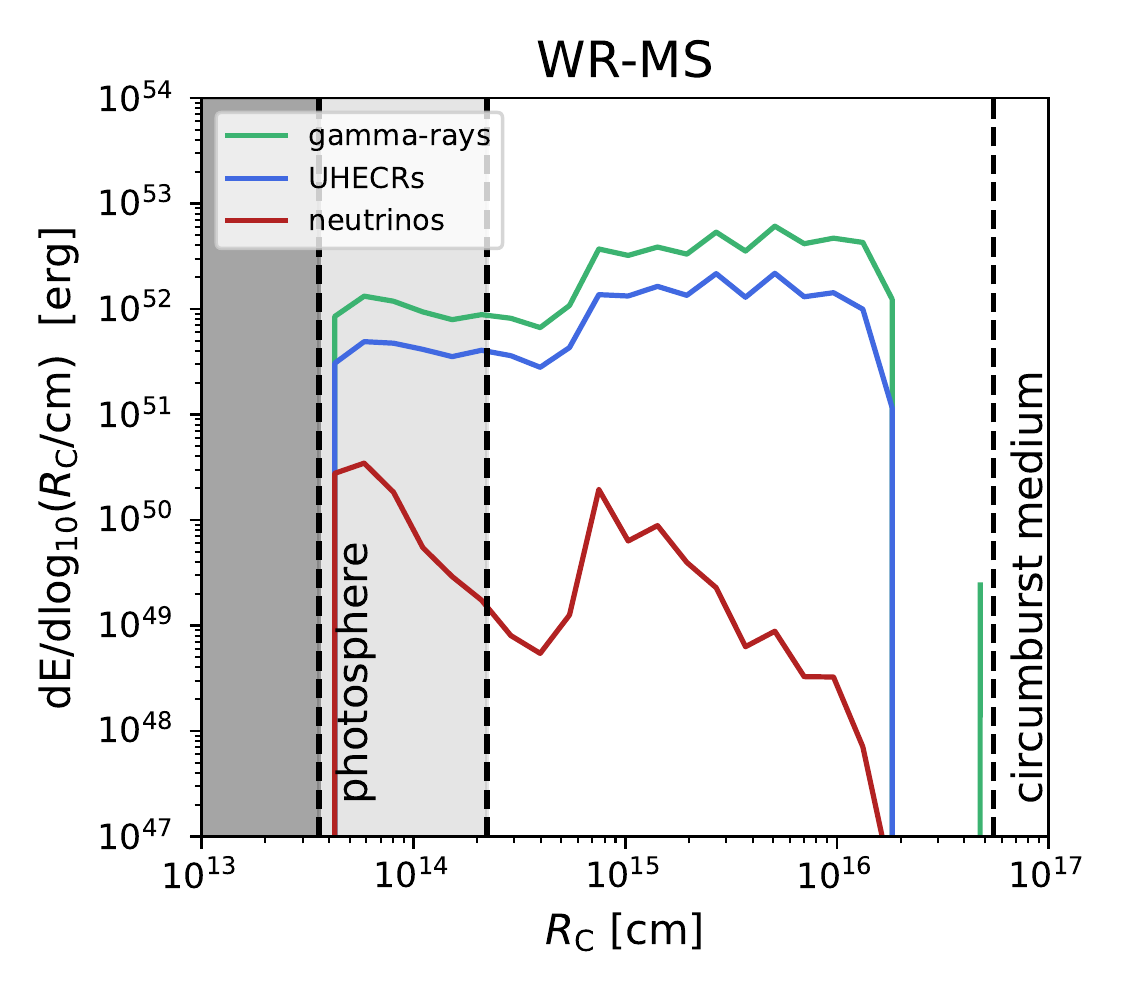}
	\includegraphics[width=.44\textwidth]{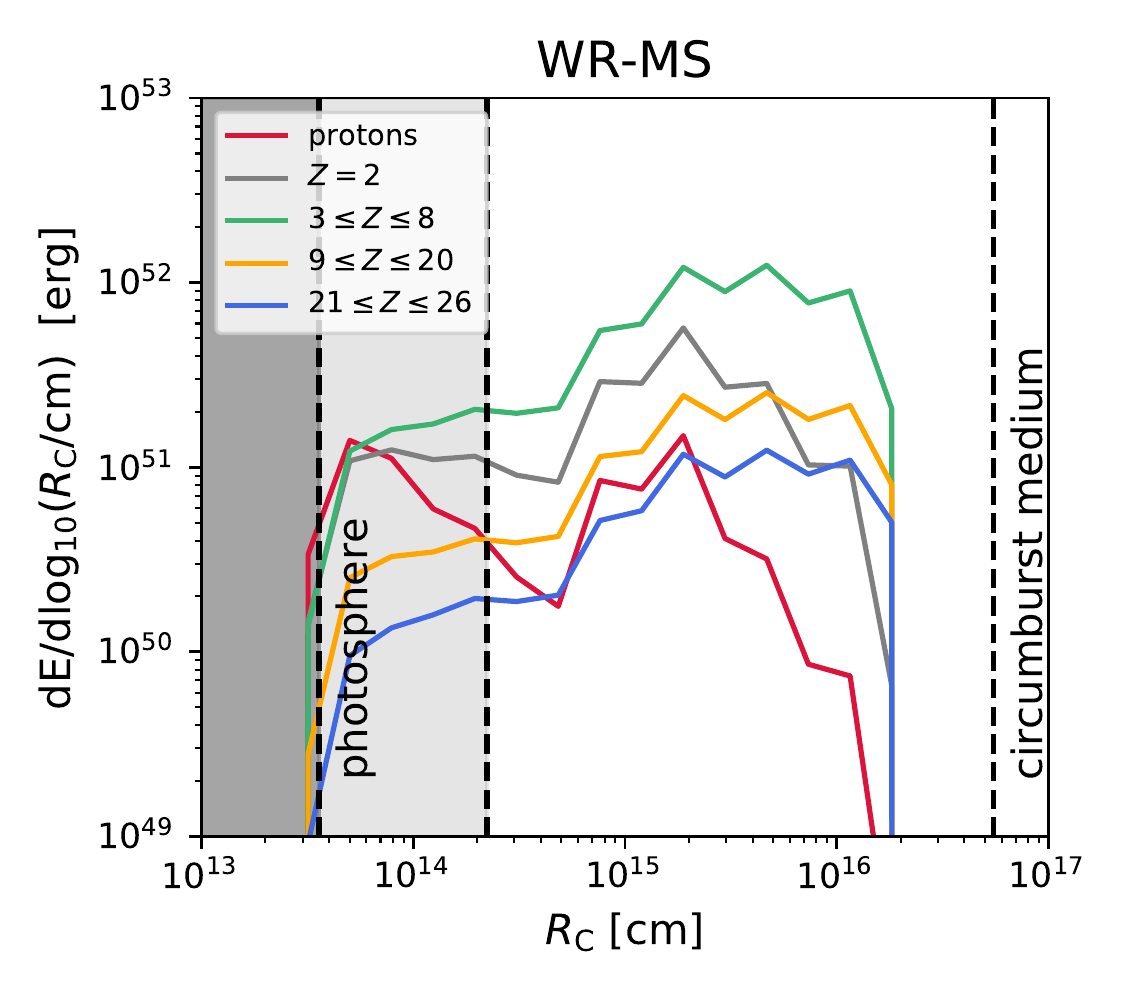}
\caption{\label{fig:particle_radii_third} Same as \fig~\ref{fig:particle_radii_first} but for \third{}  with the stochasticity parameter $A_\Gamma = 0.3$. Since collisions occur with larger variation of Lorentz factors along different radii, the photospheric condition can be encountered within the light-shaded area. The difference between the subphotospheric extrapolation (dashed) and regular emission curves reflects that the number such collisions within the light-shaded band is small. The second bump, visible at $R_C \sim 7\cdot10^14 \, \mathrm{cm}$ is related to the bulk collision and controlled by the gradient of the initial Lorentz factor distribution as described in the text. While the radial dependence at the bulk collision resembles the behavior of the \first{} case in \fig~\ref{fig:particle_radii_first}, the fireball develops efficient particle emission at much larger radii.}
\end{figure*}

Here, we discuss the two extreme examples \first{} and \third{} to illustrate the basic behavior of the source model and the effect of the source parameters. For this purpose the injection composition is fixed to identical values for the four examples, whereas in later sections the $I_A$ become free parameters of the fit.  
The particle emission regions in the left panel of \fig~\ref{fig:particle_radii_first} reflect the distribution of collisions within the fireball and thus are a direct consequence of the initial Lorentz factor distribution. Collisions occur first where the gradient in speed is maximal since $\Delta \beta = \beta_r - \beta_s \approx 1 / 2~\Gamma_s^2 -  1 / 2~\Gamma_r^2$. The ramp-up in the Lorentz factor distribution results in the bulk of the particle emission around happening at radii above $R_C \sim 7\cdot10^14$~cm.
Due to absence of stochastic engine behavior in \first{}, collisions with subsequent particle emission occur up to a radius of $5 \cdot 10^{15}\,$~cm, which corresponds to the point where the latest emitted shell runs into the bulk of slow merged shells. Neutrino emission peaks close to the photosphere where collisions are optically thick. The radial dependence of the nuclear composition, shown in the right panel of \fig~\ref{fig:particle_radii_first}, is determined by the interplay between the dominant cooling process and maximal energy and is different for each type of nucleus.

Emission of heavier nuclei starting from the CNO group is suppressed at lower radii due to photo-nuclear interactions, while efficient acceleration retains a high emission up to the maximal radius. H and He emission is highest at intermediate radii and drops toward outer radii since acceleration becomes less efficient. 

To some extent, the \first{} scenario is orchestrated since the initial Lorentz factor distribution is chosen such that the bulk of the contributing collisions occurs near the {\it bulk collision} at $R_C \sim 7\cdot10^14$~cm. 
By adding a spread $A_\Gamma > 0$ to the Lorentz factor distribution a two-bump shape develops (\fig~\ref{fig:particle_radii_third}) where the second bump comes from the bulk collision and the collisions at lower radii from collisions between neighboring shells due to random fluctuations in their Lorentz factors. The height of the second peak, and thus efficient UHECR emission, depends on the choice of $A_\Gamma$ and the amount of energy left over from the random collisions at lower radii or earlier times. If the stochasticity is excessively high, the energy remaining in the fireball for the bulk collision at larger radii is too low to produce a significant output of UHECR. At the same time, the bulk Lorentz factor decreases and harsher assumptions have to be made for $\Gamma_\text{min}$ and $\Gamma_\text{max}$.

\section{Fitting UHECR data}
\label{sec:fitting_to_data}
In this section, we include the effects of UHECR propagation \revised{in intergalactic space} and study the parameter space which describes UHECR data. We also discuss the requirements for the source energetics and the heavy mass fraction at injection from these results. 

\subsection{Source Population and UHECR transport models}

We adopt the cosmological distribution of GRBs derived by \citet{Wanderman:2009es} based on the analysis of \textit{Swift} GRBs:
\begin{align}
    R_\text{GRB}(z) = R_\text{GRB}(0) \cdot \begin{cases}
        (1 + z)^{2.1}
        &, z \leq 3 \\
        (1 + 3)^{2.1 + 1.4} (1 + z)^{-1.4}
        &, z > 3.
    \end{cases}
\end{align}

Due to computational constraints all GRBs are computed at an injected luminosity of $10^{53}\, \mathrm{erg/s}$ instead of using the full luminosity distribution. 
The suppression of radiation from subphotospheric collisions yields an emitted luminosity close to $10^{52.5}\, \mathrm{erg/s}$, which is close to the observed break luminosity in \citet{Liang:2006ci}.

For the extragalactic propagation we employ identical tools as in \citet{Heinze:2019jou}, \ie{} the numerical code {\it PriNCe} and the distributed fitting framework. All unstable nuclear isotopes lighter than iron including neutrons decay at injection or immediately at production, since the decay length is typically much shorter than the interaction length. The chain of decay products is followed down to protons, stable nuclei and neutrinos, which are produced in charged pion and muon decays. The model for the extragalactic background light (EBL) is~\citet{Gilmore:2011ks} and the photo-nuclear disintegration model is {\sc Talys} \citep{Koning:2007}. It is noteworthy that in this study, the nuclear disintegration within the source and during propagation have been simulated with the same nuclear disintegration model.
\revised{The effects of propagation in the host galaxy are not taken into account, due to the low photon fields and the negligible size of the galaxy for UHECR treated in the ballistic approximation.}

\subsection{UHECR interactions and fit method}

Each GRB model (defined by a triple $\Gamma_{\rm min}$, $\Gamma_{\rm max}$ and $A_\Gamma$) produces an individual set of nuclear and neutrino spectra at Earth for each of the five injection masses. The superposition of the nuclear spectra yields $\langle \ln{A} \rangle$ and $\sigma(\ln{A})$ that are converted into the corresponding $\langle X_\mathrm{max} \rangle$ and $\sigma(X_\mathrm{max})$ following \citet{Abreu:2013env}. 

The purpose of our study is to perform a systematic parameter space scan over the model parameters and composition. Unfortunately the number of free parameters in the model (initial distribution of Lorentz factors $\Gamma_k$, kinetic energies $E_k$, width $l_k$, and separation $d_k$ of the  shells) exceeds the constraints from the (integrated and angular-averaged) spectrum and composition measurements of UHECRs. We make an initial guess for the parameters described in \Sec~\ref{sec:GRB_model}, in particular for the initial shell distribution and injections compositions, that appear suitable to fit the spectrum. The assumption of equal energy per emitted shell imposes constrains on multiple shell parameters such as separation and width. Another criterion for a suitable model is a sufficient maximal rigidity similar to that observed in the UHECR spectrum. It results in constraints on the average emission radius and magnetic fields, which are mainly controlled by the initial distribution of Lorentz factors that translate into shell speeds.

The \textsc{Minuit}\footnote{We use the \textsc{iMinuit} interface \url{https://github.com/iminuit/iminuit}.} fitter \citep{1975CoPhC..10..343J} is used for the minimization of the goodness of fit estimator
\begin{align}
	\label{eq:chi2_definition}
    \chi^2 &= \sum_i \frac{(\mathcal{F}(E_i) - \mathcal{F}_\text{model}(E_i,\delta_E))^2}{\sigma_i^2}.
\end{align} 
The total $\chi^2$ includes individual contribution from the all-particle spectrum and the $\langle X_{\mathrm{max}} \rangle$. The data are taken from the Pierre Auger Observatory \citep{Fenu:2017,Bellido:2017} and include a systematic energy-scale uncertainty $\delta_E = \pm14\%$.
Note that we do not include  the second moment of the $X_{\mathrm{max}}$ distribution in the total $\chi^2$. This is not a technical limitation and the reasoning behind this choice is explained later. 

In order to optimize the computation, individual simulations are performed for each injection isotope separately, to be superimposed later, \ie{} in total we compute $N_{\Gamma_\text{min}} \times N_{\Gamma_\text{max}} \times N_{A_\Gamma} \times 5~\text{injection masses}=12 \times 10 \times 11 \times 5 = 6600$ fireball evolutions.
We minimize the five injection fractions $I_A$ and the energy shift $\delta_E$ of the Pierre Auger Observatory data. Note that while in the propagation computation the relation between the injected and ejected mass fraction is linear, this is only approximately true for the source model since a higher baryonic fraction may shift some collisions into the photosphere and change the emission spectra, see discussion in \citet{Bustamante:2016wpu}. We verified in separate simulations that the impact of this non-linearity is small within our current parameter ranges.

The confidence contours are drawn using $\Delta \chi^2 = \chi^2 - \chi^2_\text{min}$ projected onto planes of two parameters by minimizing over all other fit parameters. The best fit point is found by minimizing over all points in the ($\Gamma_\text{min}$, $\Gamma_\text{max}$, $A_\Gamma$) cube. 

\subsection{Systematic parameter space search}

As outlined in \Sec~\ref{sec:mm_regions} the engine parameters $\Gamma_\text{min}$, $\Gamma_\text{max}$ and $A_\Gamma$ affect the distribution of collisions along the jet and their properties, and thus impact the particle interactions and shift the emission regions of messengers.
We systematically scan the engine parameter space in $\Gamma_\mathrm{min}$, $\Gamma_\mathrm{max}$ and $A_\Gamma$ assuming that the model represents the emission of all GRB that power the UHECR flux. We continuously adjust the fraction that goes into non-thermal baryons $f_b$ and the integral injection fractions for the nuclear species such that the observed UHECR spectrum and composition is fitted for each engine configuration. As we will see, the fit contours enclose a relatively wide range of GRB realizations, implying that it is possible to obtain similar fits but with a superposition of multiple GRB models. We do not attempt this kind of generalization since it simply increases the number of free parameters.

\begin{figure*}
	\includegraphics[width=.6\textwidth]{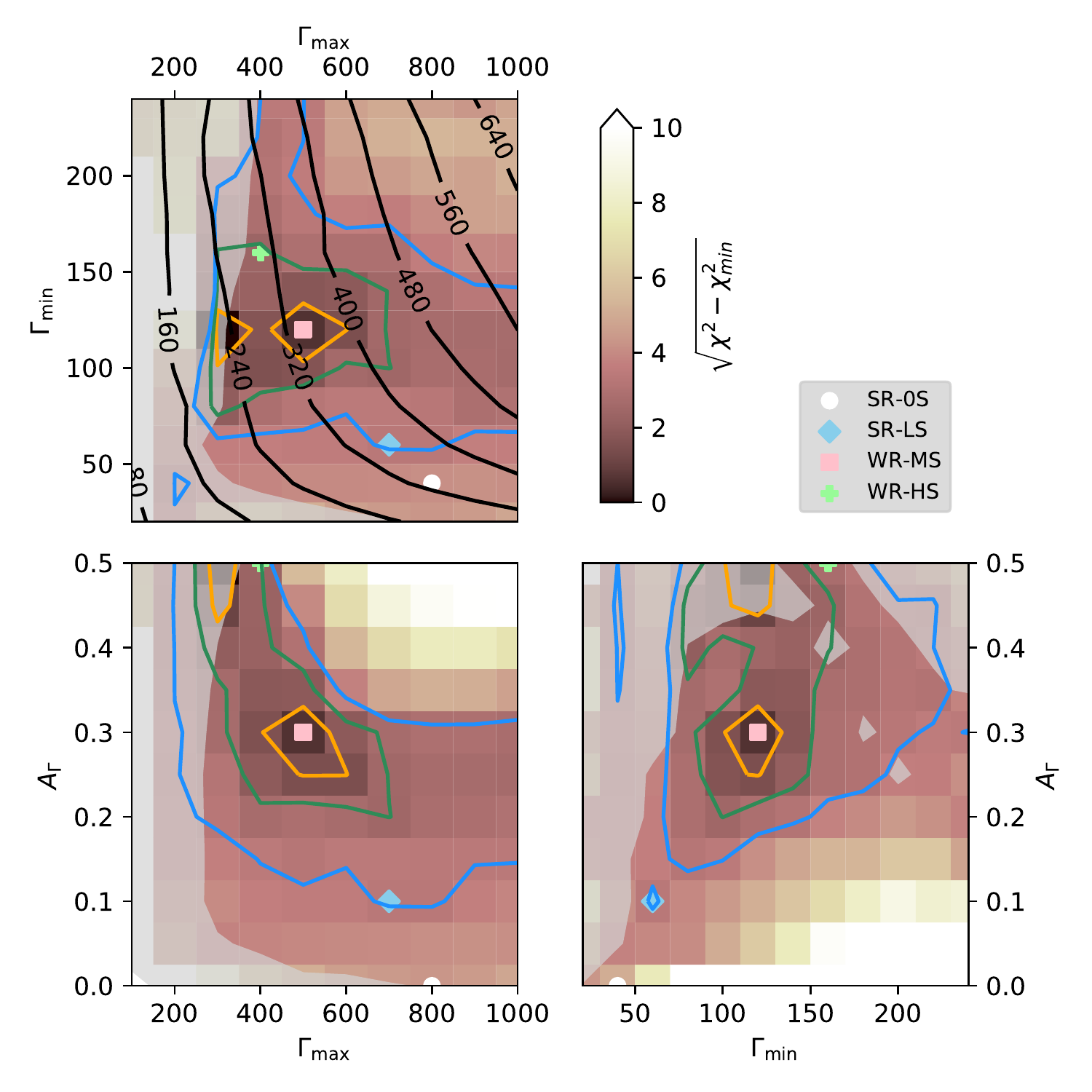}
\caption{\label{fig:paramspace} Parameter space in $\Gamma_\mathrm{max}$, $\Gamma_\mathrm{max}$ and $A_\Gamma$ for the fit to the UHECR spectrum and composition observable $\langle X_\mathrm{max} \rangle$. The orange, green and blue contours are given for $\Delta \chi^2=1$, $4$ and $9$, respectively, corresponding to $1$, $2$ and $3\sigma$ for $1$ d.o.f. for a Gaussian likelihood; see also color scale. For each two dimensional panel, the $\Delta \chi^2$ is minimized over the third (unshown) parameter and the injected composition. In the gray shade area, less than $40\%$ of the energy is dissipated in super-photospheric collisions. The black contours in the upper left panel correspond to the bulk Lorentz factor of the initial shell setup. Our benchmark scenarios are marked as in the figure legend. While the actual best-fit is in the left orange contour in the upper left panel, we use \third\ as similar scenario which has a $\Delta \chi^2$ very close-by (and a slightly higher bulk Lorentz factor which is comparable to the other benchmarks).
}
\end{figure*}

The fit result is shown in \fig~\ref{fig:paramspace} where the third variable is integrated out or summed over for each two-dimensional distribution. For each point in $\Gamma_\text{min}$, $\Gamma_\text{max}$ and $A_\Gamma$, the five baryonic loadings, the energy scale uncertainty of the data has been obtained by (continuous) minimization. The contours for the 3$\sigma$ confidence interval enclose a large parameter space, demonstrating that the model (fit) is relatively robust against parameter changes. Engines without stochastic behavior $A_\Gamma = 0$ are disfavored as well as those with $\Gamma_\text{max} < 200$. The 1- and 2$\sigma$ contours favor engines that produce an average $\Gamma_\text{bulk} \sim 200 - 400$ indicated by the iso-contours in the upper left panel. 
That is {\em per se} interesting, as a bulk Lorentz factor around 300 has been reconstructed from observations (see, for example, \citet{Ghirlanda:2017opl}), which we recover by completely different means.
The best fit belongs to the \third{} engine type with a weak ramp up and medium stochasticity. Our model is not applicable within the white shaded region at lower $\Gamma_\text{min}$ in which more than 60\% of energy is dissipated below the photosphere. Except a small overlap with the 1-$\sigma$ contour and a narrower secondary minimum with a very high stochasticity, the UHECR fit prefers GRB realizations for which most energy is dissipated above the photosphere. 

For the discussions that follow further below, we choose four examples indicated by the markers in \fig~\ref{fig:paramspace}. The rough criteria to choose these examples we motivated by a similar $\Gamma_\text{bulk} \sim 320$ as the best fit (\third{}), very different stochasticity (\first{} vs \fourth{}) and one example with a low $A_\Gamma$ (\second{}). All examples lie within (or close to) the three sigma contours. 

The details on the fit result are summarized in \Tab~\ref{tab:examples}, which we will also come back to later. The baryonic loadings are comparable to the simplistic expectation if equal number rates of electrons and protons are picked up at similar (low energies) by the acceleration process, for which one obtains $f_b \simeq (m_p/m_e)^{1/2} \simeq 44$, see \eg\ \citet{1993A&A...270...91P}. The dissipation efficiency is around 20\%, which is within the typical range expected for GRB internal shock scenarios, see \citet{Rudolph:2019ccl} for a more detailed discussion. 

\begin{table*}
\caption{Parameters for the four benchmark cases; see main text for details. 
\label{tab:examples}}
	\begin{tabular}{l|cccc}
	   \hline
		& \first   & \second  & \third & \fourth  \\\hline
		$\Gamma_\text{max}$
		& 800 & 700 & 500 & 400  \\
		$\Gamma_\text{min}$
		& 40 & 60 & 120 & 160  \\
		$A_\Gamma$
		& 0.0 & 0.1 & 0.3 & 0.5 \\
		$\chi^2$
		& 51.0 & 34.3 & 23.4 & 30.7 \\
		$\chi^2/\mathrm{dof}$
		& 3.9 & 2.6 & 1.8 & 2.4 \\
		\hline\hline
		Baryonic loading $f_b$
		& 80.1 & 67.1 & 59.5 & 108.4 \\
		Energy shift $\delta_E$
		& 0.14 & -0.14 & -0.14 & -0.14 \\
		Dissipation efficiency $\epsilon_{\mathrm{diss}}$
		& 0.28 & 0.22 & 0.13 & 0.14  \\
		Fraction super-photospheric $f_{\mathrm{sup}}$
		& 0.67 & 0.80 & 0.82 & 0.43\\
		\hline
		$E_\gamma$
		& 6.67$\cdot 10^{52}$ erg & 8.00$\cdot 10^{52}$ erg & 8.21$\cdot 10^{52}$ erg & 4.27$\cdot 10^{52}$ erg  \\
		$E_{\mathrm{UHECR}}^{\mathrm{esc}}$ (escape)
		& 2.01$\cdot 10^{53}$ erg & 2.10$\cdot 10^{53}$ erg & 1.85$\cdot 10^{53}$ erg & 1.69$\cdot 10^{53}$ erg  \\
		$E_{\mathrm{CR}}^{\mathrm{src}}$ (in-source)
		& 5.11$\cdot 10^{54}$ erg & 5.13$\cdot 10^{54}$ erg & 4.62$\cdot 10^{54}$ erg & 4.36$\cdot 10^{54}$ erg  \\
		$E_{\mathrm{UHECR}}^{\mathrm{src}}$ (in-source, UHECR)
		& 3.70$\cdot 10^{53}$ erg & 4.46$\cdot 10^{53}$ erg & 3.97$\cdot 10^{53}$ erg & 3.57$\cdot 10^{53}$ erg  \\
		$E_\nu$
		& 7.81$\cdot 10^{49}$ erg & 2.18$\cdot 10^{50}$ erg & 1.28$\cdot 10^{51}$ erg & 1.79$\cdot 10^{51}$ erg  \\
		$E_\mathrm{kin,init}$
		& 2.90$\cdot 10^{55}$ erg & 3.03$\cdot 10^{55}$ erg & 4.50$\cdot 10^{55}$ erg & 7.81$\cdot 10^{55}$ erg  \\
		\hline
		Fraction $I_{\mathrm{H}}$
		& $0.22_{-0.05}^{+0.04}$ & $0.00_{-0.00}^{+0.10}$ & $0.00_{-0.00}^{+0.06}$ & $0.01_{-0.01}^{+0.07}$ \\
		Fraction $I_{\mathrm{He}}$
		& $0.00_{-0.00}^{+0.01}$ & $0.07_{-0.05}^{+0.04}$ & $0.07_{-0.07}^{+0.07}$ & $0.27_{-0.05}^{+0.05}$ \\
		Fraction $I_{\mathrm{N}}$
		& $0.39_{-0.04}^{+0.04}$ & $0.29_{-0.08}^{+0.06}$ & $0.13_{-0.13}^{+0.11}$ & $0.00_{-0.00}^{+0.09}$ \\
		Fraction $I_{\mathrm{Si}}$
		& $0.33_{-0.04}^{+0.03}$ & $0.63_{-0.03}^{+0.03}$ & $0.76_{-0.03}^{+0.03}$ & $0.53_{-0.03}^{+0.03}$ \\
		Fraction $I_{\mathrm{Fe}}$
		& $0.06_{-0.02}^{+0.02}$ & $0.01_{-0.01}^{+0.02}$ & $0.05_{-0.03}^{+0.03}$ & $0.19_{-0.03}^{+0.02}$  \\
		\hline
		Heavy mass fraction
		& $0.78_{-0.10}^{+0.22}$ & $0.93_{-0.13}^{+0.07}$ & $0.93_{-0.19}^{+0.07}$ & $0.72_{-0.06}^{+0.28}$ \\		
		\hline 
	\end{tabular}
\end{table*}

\subsection{Source spectra}

\begin{figure*}
	\begin{center}
		\includegraphics[width=.4\textwidth]{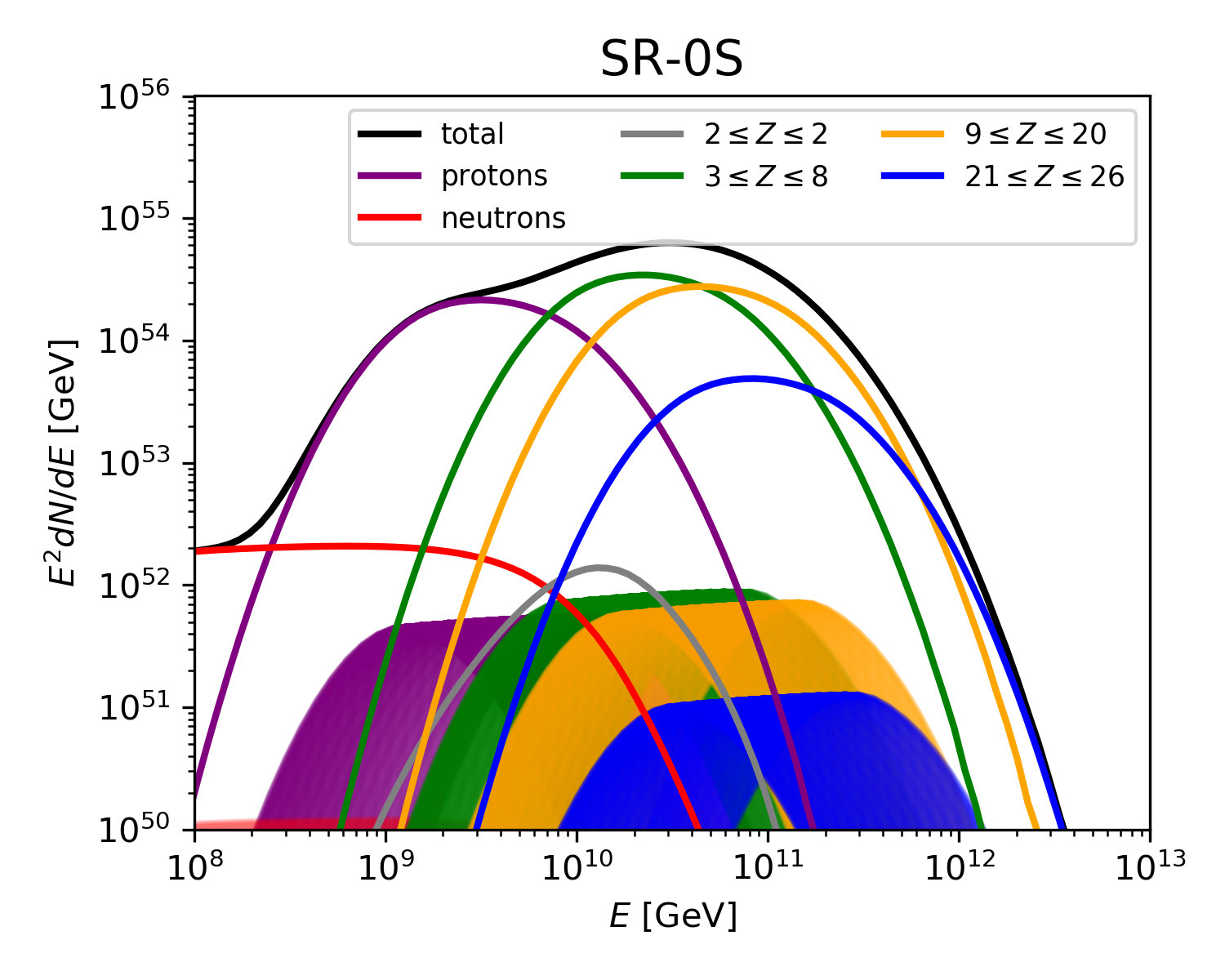} \hspace*{0.05cm}
		\includegraphics[width=.4\textwidth]{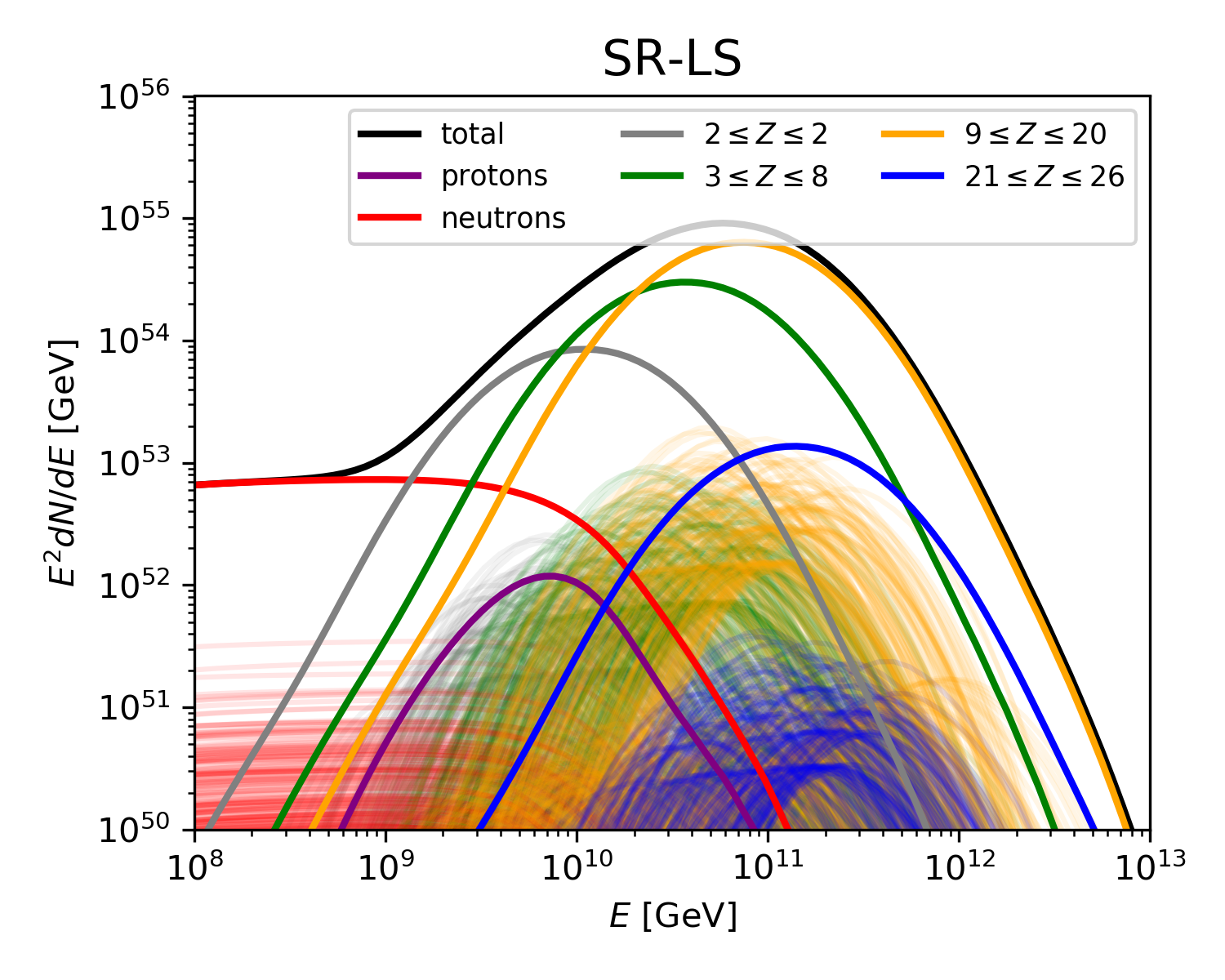} 
		
		\includegraphics[width=.4\textwidth]{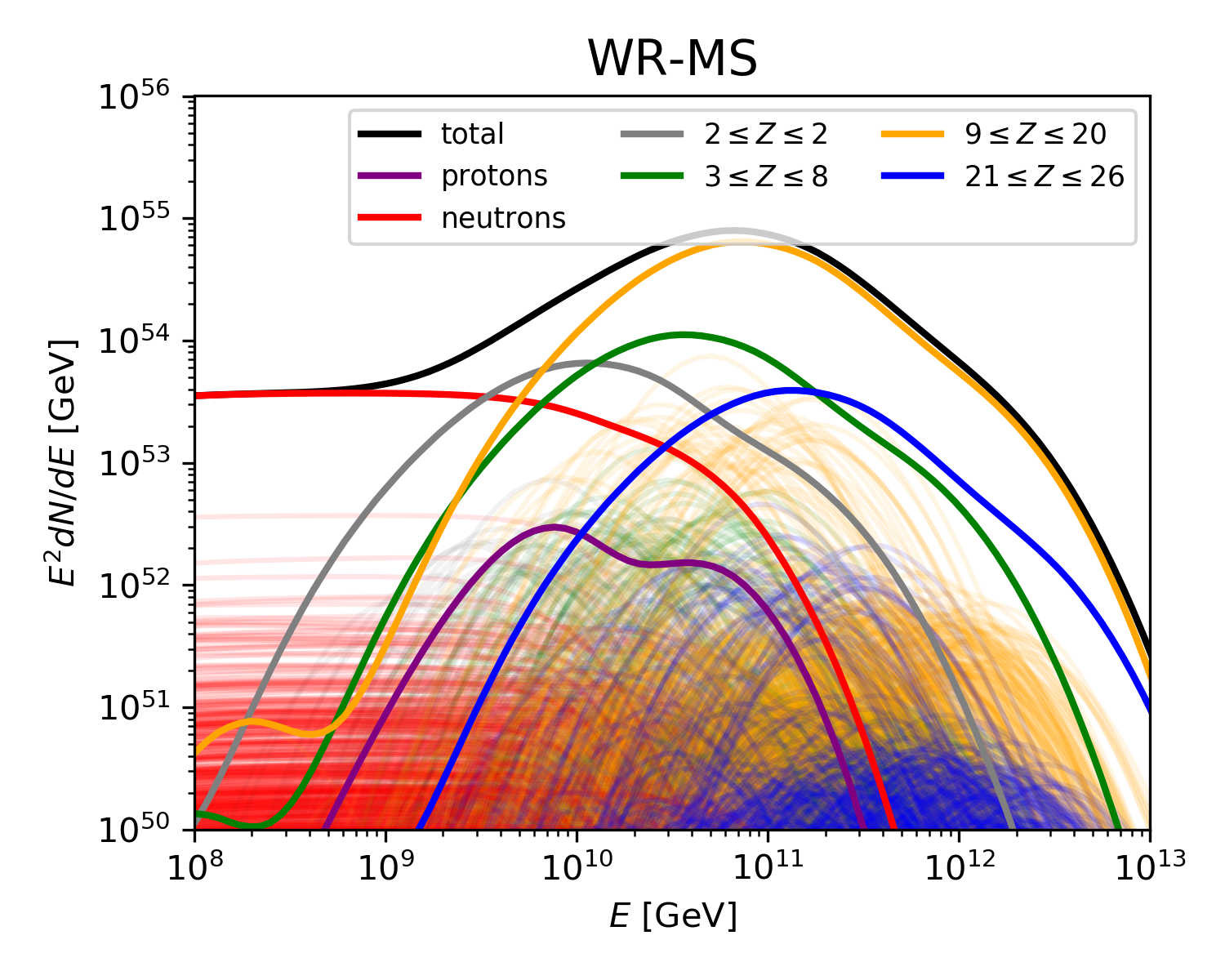} \hspace*{0.05cm}
		\includegraphics[width=.4\textwidth]{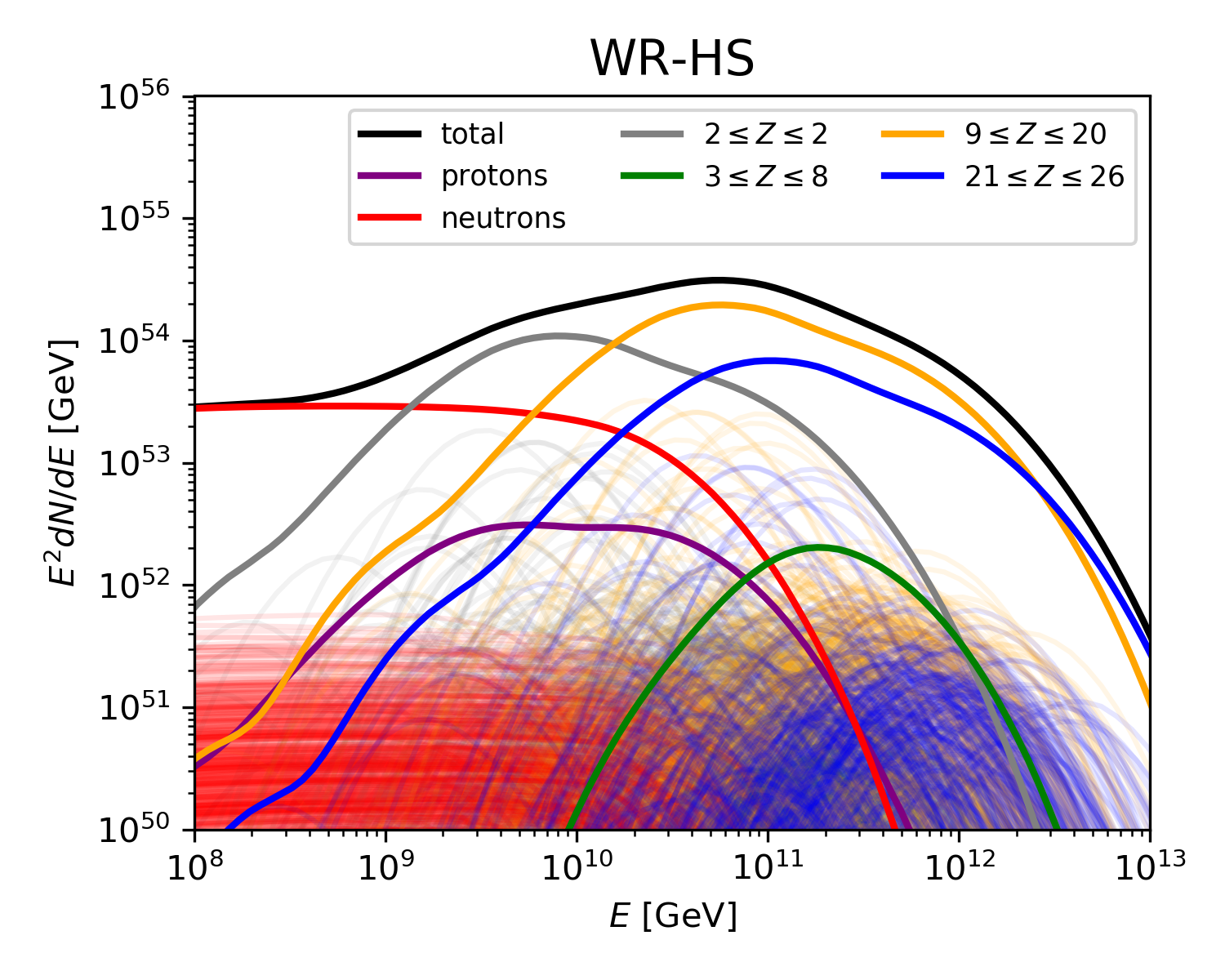}
	\end{center}
	\caption{\label{fig:source_spectra}Particle spectra ejected from the source for the four example cases. \revised{The total spectrum is shown as thick black solid curve, whereas thick colored curves show the total spectrum of each charge grouped according to the legend.} 
	 The thin curves indicate the ejected spectrum per individual collision and mass group\revised{, shaded areas are created by overlap of many such individual collisions.} }
\end{figure*}

To illustrate how different engine behaviors affect the ejected cosmic-ray spectra, we show in \figu{source_spectra} the ejected spectrum from the source for each of the four example cases. The weights for the five intregral fractions of the injection masses are obtained from the fit, the initial composition of the jet is thus different for each of the example cases. 

By adding stochasticity the emission of nuclei is spread out among a larger range of radii compared to the case of no stochasticity (\first). With stochasticity, collisions can occur at smaller radii, where optical depths for photo-nuclear interactions are high and more secondary nucleons (red curves) are produced through disintegration. This also leads to higher neutrino emission. 

The smaller cosmic-ray production region close to the engine of \first{} is also reflected in smaller maximal energies of cosmic-ray nuclei, as the upper left panel of \figu{source_spectra} clearly shows. Higher maximal energies are reached by stochastic engines through collisions that occur at outer radii where the acceleration of heavier nuclei is still efficient but the source is transparent enough for nuclei to escape. These collisions will negligibly contribute to the total neutrino output. The distribution of collisions among many radii in stochastic models can, however, yield neutrino bright and UHECR dim, and, UHECR bright and neutrino dim collisions within the same astrophysical object in contrast to one-zone models as in \citet{Biehl:2017zlw} or multi-collision models dominated by one bulk collision (\first).

Concerning the spectral shape, the contributions of collisions that are spread out along all radii in models with higher stochastic component lead to a softening/broadening of the emission spectra for individual mass groups. This is demonstrated in \figu{source_spectra} where the spectra from each collision are illustrated by thin and the total output by thick curves. For models dominated by the \textit{bulk collision} (\first{} and \second{}) the spectra are clearly narrower. 

\subsection{UHECR spectra at Earth and general aspects of the fit}
\begin{figure*}
	\includegraphics[width=.44\textwidth]{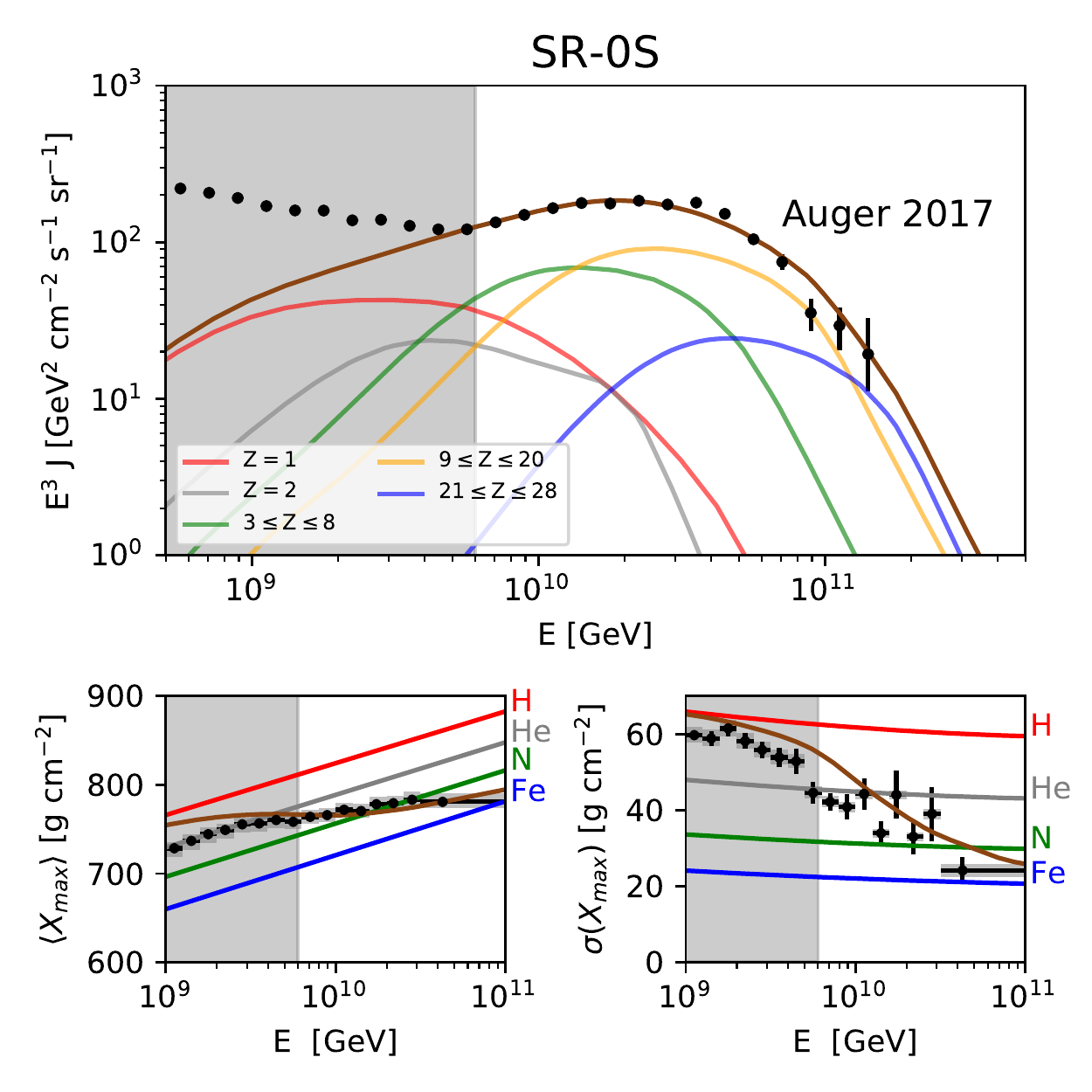}
	\includegraphics[width=.44\textwidth]{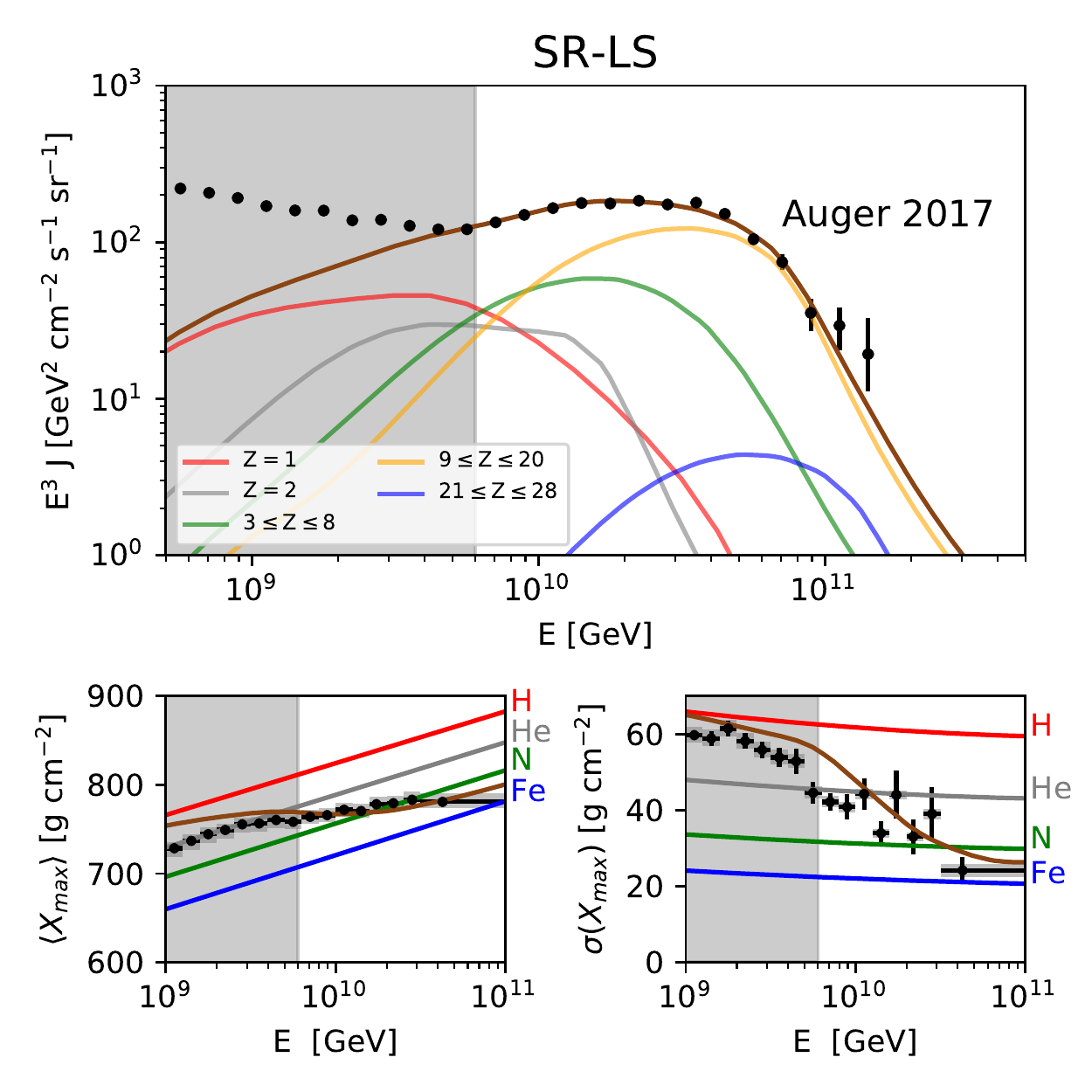}
	\includegraphics[width=.44\textwidth]{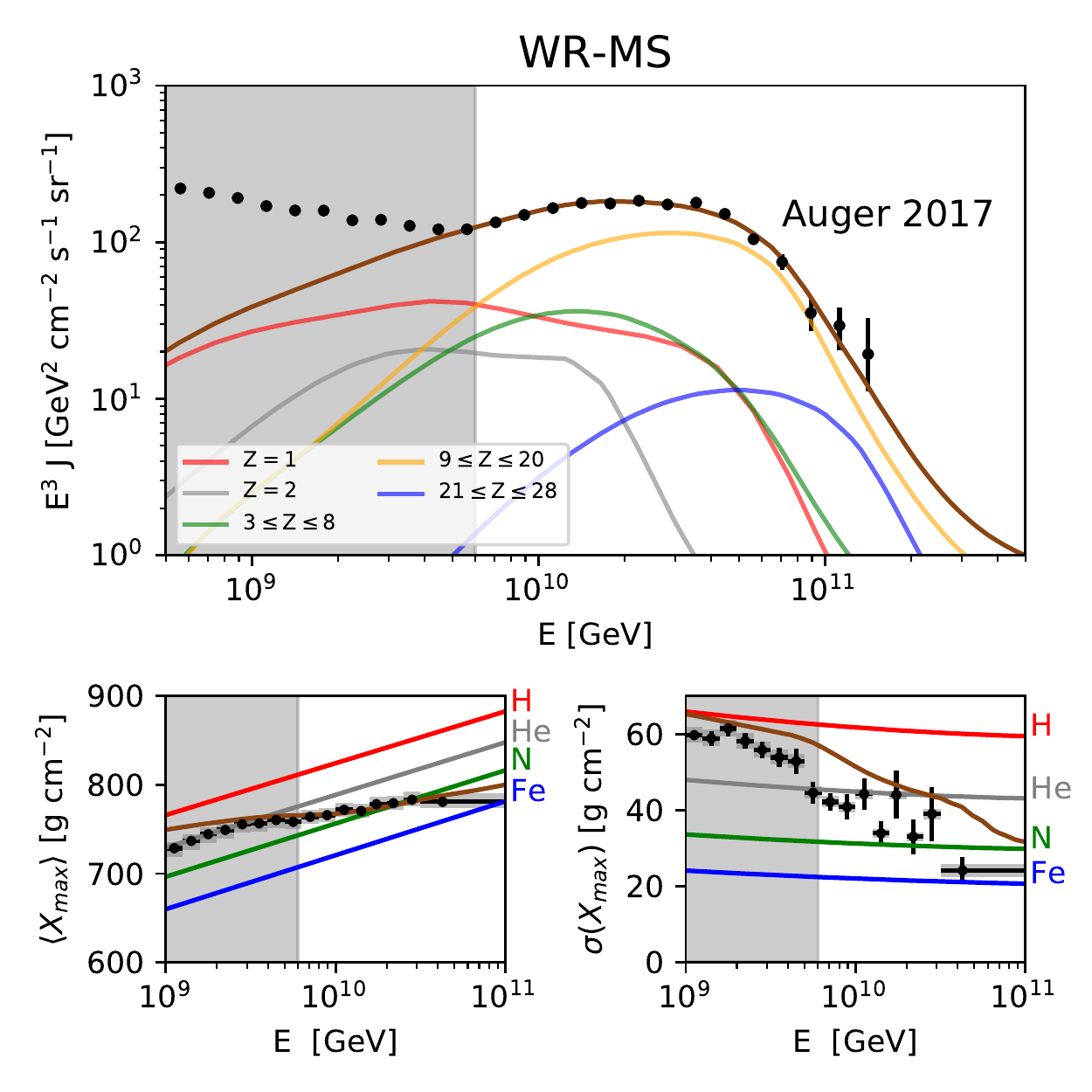}
	\includegraphics[width=.44\textwidth]{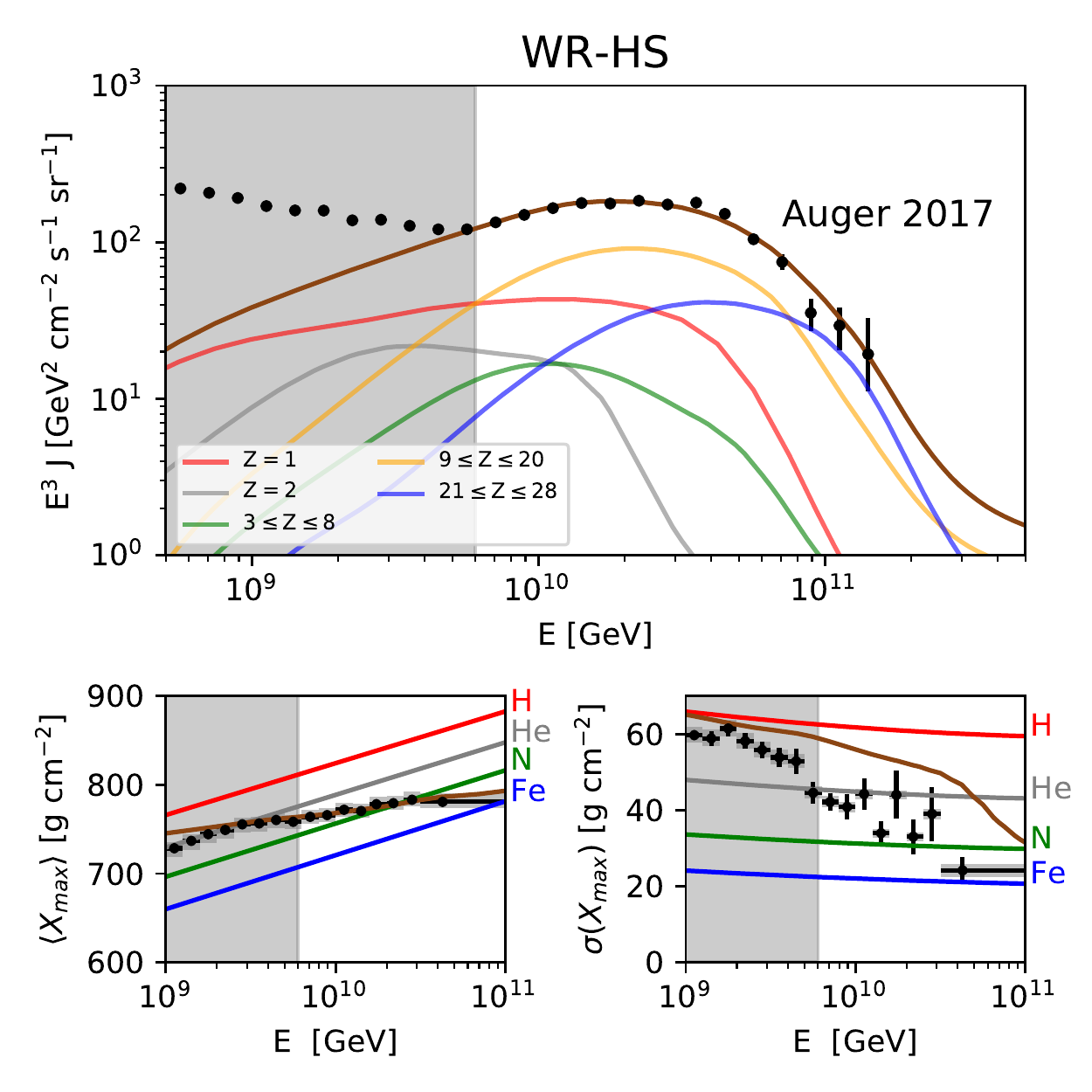}
\caption{\label{fig:observed_spectra} Observed UHECR spectrum (large panels) for two moments of $X_\mathrm{max}$ (small panels) for the four example cases for the best-fit composition (see \Tab~\ref{tab:examples}). Only the spectrum and mean $\langle X_\mathrm{max} \rangle$ are included in the fitting procedure. The gray shaded area indicates the range below $6 \cdot 10^{10}~\mathrm{GeV}$ which is excluded from the fit.}
\end{figure*}

Following the ejected spectra we discuss the corresponding observed spectra and general properties of the fit for the four example cases.

Comparing observed to ejected spectra, one has to bear in mind general effects of extragalactic propagation. Interactions with background photons during propagation generally lead to a flux depletion of nuclei with energies above the threshold for photo-disintegration. Secondary nuclei that emerge from photo-nuclear interactions populate the spectra at lower energies, softening the spectrum at Earth compared to that at the source. Because the boost is approximately conserved during disintegration, the energy $E_p$ of the interacting nucleus of mass $A_p$ and a secondary nucleus of mass $A_s$ is fixed by the relation $E_s=A_s/A_p \cdot E_p$. The lighter secondary nuclei from disintegration, therefore, show up at lower energy. 

The spectra at Earth are shown in \fig~\ref{fig:observed_spectra}. It is remarkable how well the spectrum and the $\langle X_{\mathrm{max}}\rangle$ are reproduced with such different source spectra and engine properties. 

To fit the observed UHECR spectrum for different engine realisations, we adjust the integral fractions of the injection elements at the source (\eq~\ref{equ:frac}) for the four benchmark cases. The integral fractions resulting from the fit are listed in Table~\ref{tab:examples}.
In all cases data require more than 50\% of intermediate nuclei (N and Si) injected at the source. The \third{} and \fourth{} cases require an even higher ($>70\%$) integral fraction of silicon and iron at injection to compensate for their depletion due to efficient photo-disintegration in the source. An unavoidable counterpart to efficient photo-disintegration is the high abundance of secondary neutrons in the source (thin red curves in \fig~\ref{fig:source_spectra}). This is reflected in the integral fractions that do not require any additional proton injection in the source when stochasticity is present.

In the absence of stochasticity the \first{} case requires a 22\% proton integral fraction at injection to compensate for the low disintegration in the source. Since the high maximal energies required by data are barely reached at the source, the fit pulls the energy scale ($\delta_E < 0$ means data move toward lower energy) to minimize the tension. Because maximal energies are smaller, the missing abundance of light elements is not well compensated through disintegration during propagation. The high maximal energies at the source for \third{} and \fourth{} (visible in the lower panels of \fig~\ref{fig:source_spectra}) affect the spectrum at and beyond the cutoff energies in the lower panels of \fig~\ref{fig:observed_spectra} that show an onset of recovery above $2 \cdot 10^{11}$ GeV.

Interestingly, the \first{} case does not require any (primary) helium to be injected into the source. Therefore, all helium is secondary, \ie{} a product of photo-disintegration in the source or the propagation. However, we notice that this might also be an effect of the disintegration model (\textsc{Talys}) used in the computation. The production of helium, and its subsequent disintegration, is strongly affected by the uncertainties in the disintegration cascade, due to absence of data and models \citep{Batista:2015mea,Boncioli:2016lkt}. For the stochastic cases the origin of helium in UHECR at Earth is both primary and secondary. 

In simpler models, the succession of increasingly heavier mass spectra toward higher energies is often assumed to be caused by a maximal rigidity reached by the accelerator or one acceleration zone. As we show with these stochastic multi-collision models, this assumption is not essential to describe the spectrum and the $\langle X_{\mathrm{max}}\rangle$. In \second{} and \third{} the heaviest mass group at the cutoff is mostly silicon and not iron. For \third{} and \fourth{} the cutoff for the proton spectrum at Earth reaches or exceeds that of the helium or the nitrogen group, confirming that the data do not require that the maximal energy at ejection follows the Peters cycle, as already found for example in \citet{Biehl:2017zlw}. This is already visible in the source spectra and is enhanced at Earth since protons are abundantly produced during propagation of heavier elements with energies close to, or above the observed cutoff.

A real discrimination of the models and a crucial piece of the origin of UHECR puzzle may come from the width of the $X_\mathrm{max}$ distribution, $\sigma(X_\mathrm{max})$. The data requires that while the composition has to become heavier toward higher energies, this transition has to happen through a smooth gradual succession between neighboring mass groups (see \eg{} \cite{Aab:2016zth}) ,which minimizes the overlap of different mass spectra.
For example, the $\langle X_{\mathrm{max}}\rangle$ could in principle be described by a mixture of just protons and iron. The overlap of protons and heavier spectra would, however, lead to large fluctuations in $X_{\mathrm{max}}$ producing $\sigma(X_\mathrm{max})$ distribution that is flat or increasing in energy, contrary to what is observed.

Clearly, our model does not describe $\sigma(X_\mathrm{max})$ sufficiently well. If we include $\sigma(X_\mathrm{max})$ in the $\chi^2$ definition in \eq~(\ref{eq:chi2_definition}), the result only marginally improves at the cost of narrow contours. The preferred models tend to have fewer optically thin collisions contributing to the emission and mostly lie at the edge of the range in which our model is applicable. The reasons for this behavior are twofold and can be understood by comparing \figs~\ref{fig:source_spectra} and \ref{fig:observed_spectra}:
\begin{enumerate}
	\item Since collisions from different radii contribute to the total output, the maximal energies of individual collisions cover a wider range (see the colored ``bands'' in the \first{} panel of \fig~\ref{fig:source_spectra} that are comprised of successive thin curves). When those are summed (thick, colored curves for each mass group) the total spectra are much wider than the individual bell shapes, equivalent to a softening of the mass group spectra. For stochastic engines this effect is stronger, since there is a wider spread in the maximal energy of individual positions. In combination with a positive source evolution, softer spectra are known to produce worse fits.
	\citet{Heinze:2019jou} showed that this behavior persists for positive redshift evolutions across various model combinations used for the simulation of air showers and extragalactic propagation.
	\item The second reason is the strong overlap of light and heavy mass spectra in stochastic models, in particular the abundance of protons at the highest energies in the lower panels of \fig~\ref{fig:observed_spectra}. The required smooth succession of heavier masses is violated and hence $\sigma(X_\mathrm{max})$ can not decrease as in the data. This is related to the level of disintegration in the source, the maximal energy required to fit the observed spectrum, and the level of disintegration of heavy elements on the extragalactic photon fields during propagation. A significant suppression of in-source disintegration is not easily possible if electrons and nuclei are accelerated by the same mechanism and radiate within the same volume.   
\end{enumerate}

We note that similar effects will occur when giving up the assumption of identical GRBs. If multiple GRBs or other generic accelerators with different maximal rigidities or intrinsic luminosities contribute to the observed average UHECR flux, the average of source spectra will be softer/broader than that of individual objects.
The same problem in describing $\sigma(X_\mathrm{max})$ will therefore occur for any model that uses a population of distant sources with different maximal rigidities.
On the other hand, a mechanism that generally suppresses photo-disintegration may reduce the tension of the model with $\sigma(X_\mathrm{max})$ observations, but would likely also reduce emission of secondary messengers such as photons or neutrinos, rendering the attempt to discover UHECR sources with multimessenger techniques a stiffer challenge than it already is.

\subsection{Source energetics}

A well known problem for the GRB origin of UHECRs is the large required isotropic-equivalent energy emitted in baryons around $2-3 \cdot 10^{53}\,$erg per GRB in the UHECR range~\citep{Baerwald:2014zga} (here computed for the Wanderman-Piran GRB evolution). This estimate is a consequence of the required local emissivity $\sim 10^{44} \, \mathrm{erg \, Mpc^{-3} \, yr^{-1}}$ to sustain the flux of UHECRs with a local GRB rate around $1 \, \mathrm{Gpc^{-3} \, yr^{-1}}$. This implies that the required kinetic energy must be much larger, even if most of it is efficiently transferred into escaping non-thermal baryons {\em with the highest energies}. The dissipation efficiency (transfer from kinetic energy to non-thermal radiation) is only one part of the problem, the other is how to accelerate baryons to the highest energies without retaining much of it below the UHECR energy range. 

The four examples have been chosen to describe GRBs with gamma-ray energies in the ball park of a few times $10^{52} \, \mathrm{erg}$ (row $E_\gamma$ in \Tab~\ref{tab:examples}). In fact, the overall normalization is $E_\gamma \equiv 10^{53} \, \mathrm{erg}$ but sub-photospheric collisions do not contribute to this budget and hence the resulting energy output is lower. The fraction of energy emitted in super-photospheric collisions (see respective row) is $f_{\mathrm{sup}} \equiv E_\gamma/10^{53} \, \mathrm{erg}$ ranging between 40\% and 80\% for the majority of parameter combinations.

The initial kinetic energy $E_{\mathrm{kin,init}}$ is mostly converted into baryons due to relatively high baryonic loading. The energy ejected as UHECRs can be written as\footnote{There is also a small correction factor taking into account that not all injected energy (counted in $f_b$) is available for $E_{\mathrm{CR}}^{\mathrm{src}}$ because of radiation losses, which we neglect in these considerations.}
\begin{equation}
 E_{\mathrm{UHECR}}^{\mathrm{esc}} \simeq  E_{\mathrm{kin,init}} \times f_{\mathrm{sup}} \times \epsilon_{\mathrm{diss}} \times f_{\mathrm{esc}} \times f_{\mathrm{bol}} \, , 
 \label{equ:energetics}
\end{equation}
where $\epsilon_{\mathrm{diss}}$ is the fraction of kinetic energy dissipated into non-thermal radiation according to the collision model, $f_{\mathrm{bol}} \equiv  E_{\mathrm{UHECR}}^{\mathrm{src}}/E_{\mathrm{CR}}^{\mathrm{src}}$ is a bolometric correction describing how much energy is deposited into the UHECR range, and $f_{\mathrm{esc}} \equiv E_{\mathrm{UHECR}}^{\mathrm{esc}}/E_{\mathrm{UHECR}}^{\mathrm{src}}$ is the fraction of energy escaping the source. From the table, we find $\epsilon_{\mathrm{diss}} \simeq 0.13 - 0.28$, $f_{\mathrm{esc}} \simeq 0.5$, and $f_{\mathrm{bol}} \simeq 0.07-0.08$ for the chosen four examples. Apart from the known dissipation efficiency problem a large correction comes from $f_{\mathrm{bol}}$ that describes the fraction of non-thermal baryons in the UHECR energy range. Although our model implies that the baryons are picked up at low energies, such as in a thermal bath, this factor would be an order of magnitude larger if all non-thermal particles were accelerated up to UHECR energies or if the acceleration spectra were harder than $E^{-2}$.  

We find that the (isotropic-equivalent) kinetic energy of the outflow is $ \sim 10^{55} - 10^{56} \, \mathrm{erg}$ per GRB (see $E_\mathrm{kin,init}$ row in \Tab~\ref{tab:examples}) if the long-duration GRBs are the sole sources of UHECRs. Recent afterglow observations indicate that a substantial fraction of the bulk kinetic energy can escape observation~\citep{Acciari:2019dxz, Arakawa:2019cfc}, supporting earlier arguments \citep{Fan:2006sx, Beniamini:2015eaa, Beniamini:2016hzc} about a systematic underestimation of kinetic energy. But even in view of the current results our findings may sound excessive. Some ingredients of this estimate are certainly model-dependent. For example, an order of magnitude lower values $ \sim 10^{54} - 10^{55} \, \mathrm{erg}$ are found for hard (significantly harder than $E^{-2}$) acceleration spectra, or when assuming an acceleration mechanism with an extremely high transfer efficiency to the highest energies. Nevertheless, from the energetics point of view a paradigm shift is required if conventional GRBs are indeed the sources of UHECRs. Neutrino observations are an opportunity to independently test this scenario.

It is interesting to compare our result to \citet{Gottlieb:2020ifs}, who find that ``all intermittent jets are subject to heavy baryon contamination that inhibits the emission at and above the photosphere''. As we shown this baryon contamination is needed to describe UHECRs. Our radiative (not dissipation) efficiency (energy in photons versus kinetic energy) is between $5 \cdot 10^{-4}$ and $0.003$ (see \Tab~\ref{tab:examples}), which is comparable to their result. 
So perhaps their negative conclusion for the gamma-ray signal actually indicates that GRBs with intermittent engines are indeed efficient cosmic ray accelerators -- for which a large baryon contamination is required.
In contrast to the choice of generic intermittance patters in \citet{Gottlieb:2020ifs}, our empirical Lorentz factor profile and its parameters tend to suppress early sub-photospheric collisions and avoid the rapid slow down of the wind. Within our framework, we can not answer if such patterns can be realized in ``ab initio'' hydrodynamical simulations.

\subsection{Heavy mass fraction at injection}

\begin{figure*}
	\includegraphics[width=.45\textwidth]{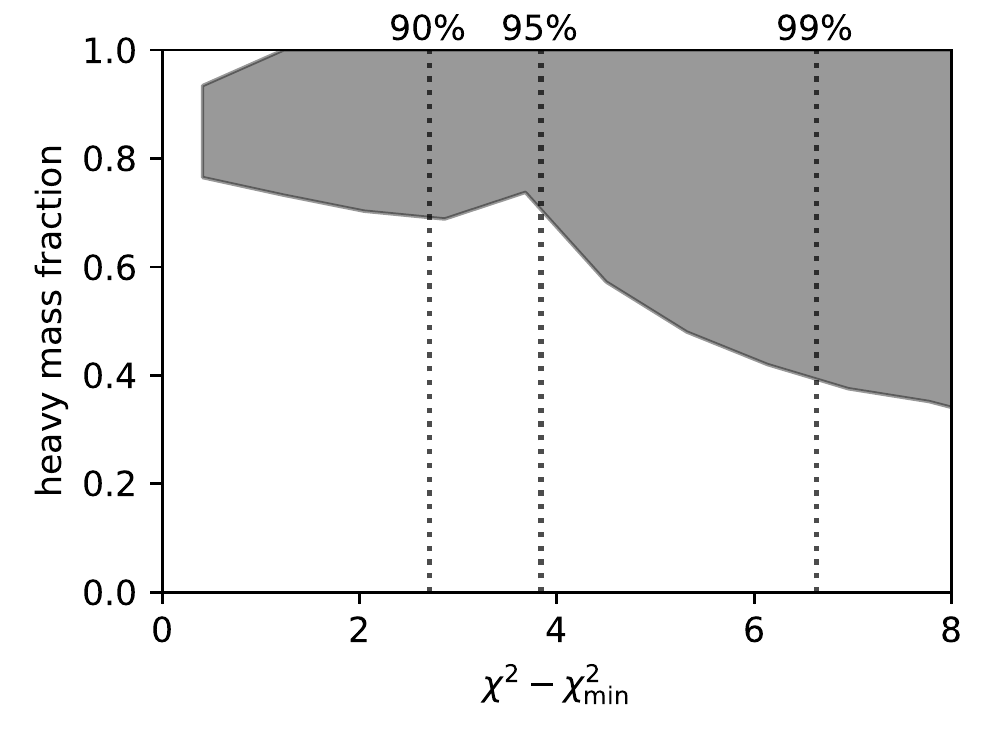}
	\includegraphics[width=.45\textwidth]{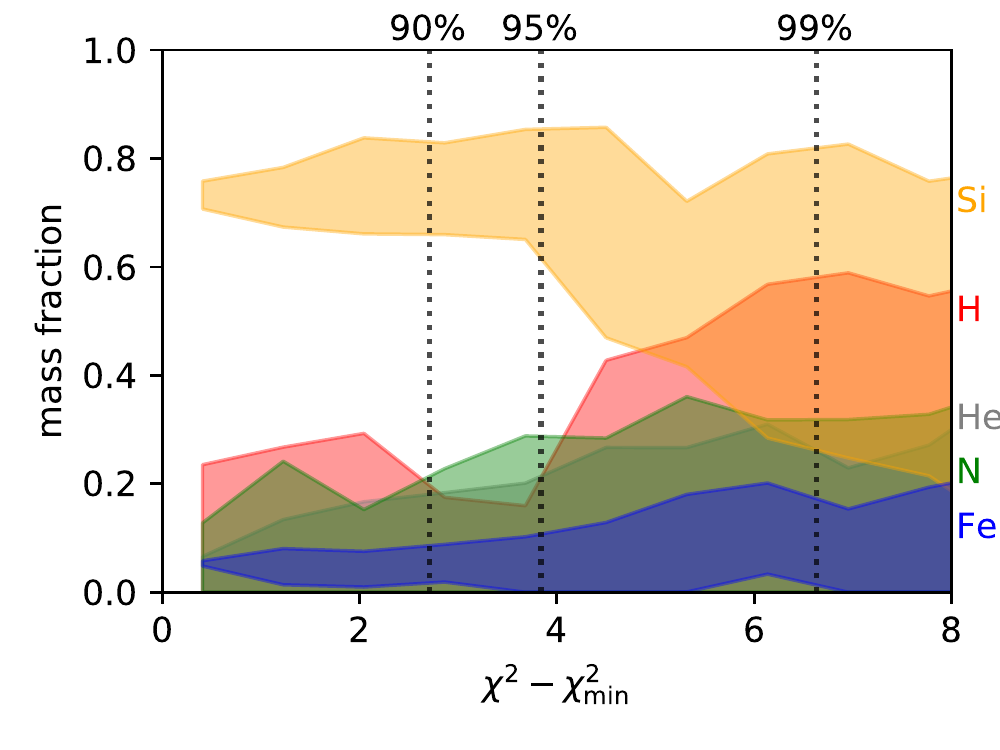}
	\caption{\label{fig:metallicity_band}Allowed heavy mass fraction (HMF, left panel) and fraction bands (right panel), both integrated over the energy and defined at injection, within the $\chi^2-\chi_{\mathrm{min}}^2$ value obtained from the UHECR fit.}
\end{figure*}

The heavy mass fraction (HMF), defined as the fraction of integrated energy released into isotopes heavier than He, is reported in the last row of \Tab\ref{tab:examples}, corresponding to the example cases and the best fit.  The increase in the stochasticity clearly enhances the efficiency of the interactions in the engine, imposing a higher injection of species heavier than He in order to fit the measured composition. The WR-HS, being extreme in terms of stochasticity, is less efficient in providing enough energy at larger radii, and the requested HMF are smaller.

The HMF is above 0.7 within the 95\% CL (see \Fig\ref{fig:metallicity_band}, left panel), and above 0.4 at the $3\sigma$ CL. This is mostly due to the increase in the proton fraction (see \Fig\ref{fig:metallicity_band}, right panel) which is allowed within the three sigma contour, that includes a smaller level of stochasticity with respect to the one sigma contour. The large value of the HMF is an outcome of the fit, compared to \cite{Globus:2014fka}, where it is ten times larger than found in Galactic cosmic rays.

The separation of heavy and light masses at injection becomes milder when the $\sigma(X_\mathrm{max})$ is included in the fit procedure. For this reason, the allowed HMF is larger than 0.6 at $3 \sigma$ if the fit includes the $\sigma(X_\mathrm{max})$, and the spread of the mass fractions is much less pronounced.

In \cite{Zhang:2017moz} several pre-supernova models have been selected from \cite{Woosley:2005gy}, to test the GRBs as sources of UHECRs. In particular, models which provide a heavy distribution of nuclear mass fractions at the onset of the core collapse are chosen in this paper, and such compositions are used as output of low-luminosity GRBs with internal shock model, that are supposed to power the UHECR flux at Earth. These have HMF greater than 0.9, which is found to be compatible with our result independent of the confidence level.

\section{Light curves and multi-messenger implications}
\label{sec:multi-messenger-results}
Since the UHECRs alone cannot discriminate among different model assumptions, we seek here for additional observables and multi-messenger signals to study the underlying model, in particular, light curves and the neutrino flux.

\subsection{Predicted light curves}

\begin{figure*}
	\includegraphics[width=.35\textwidth]{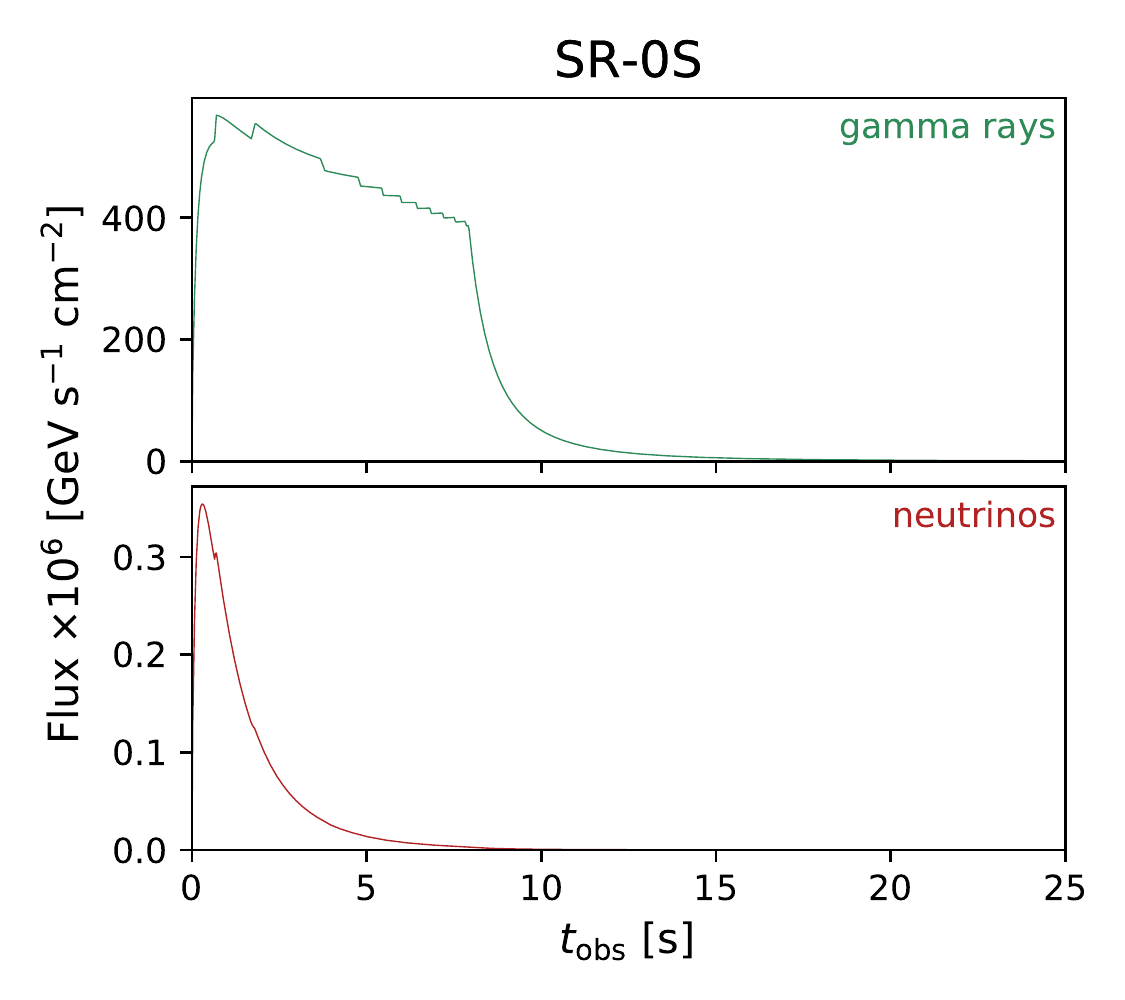} \hspace*{0.05\textwidth}
	\includegraphics[width=.35\textwidth]{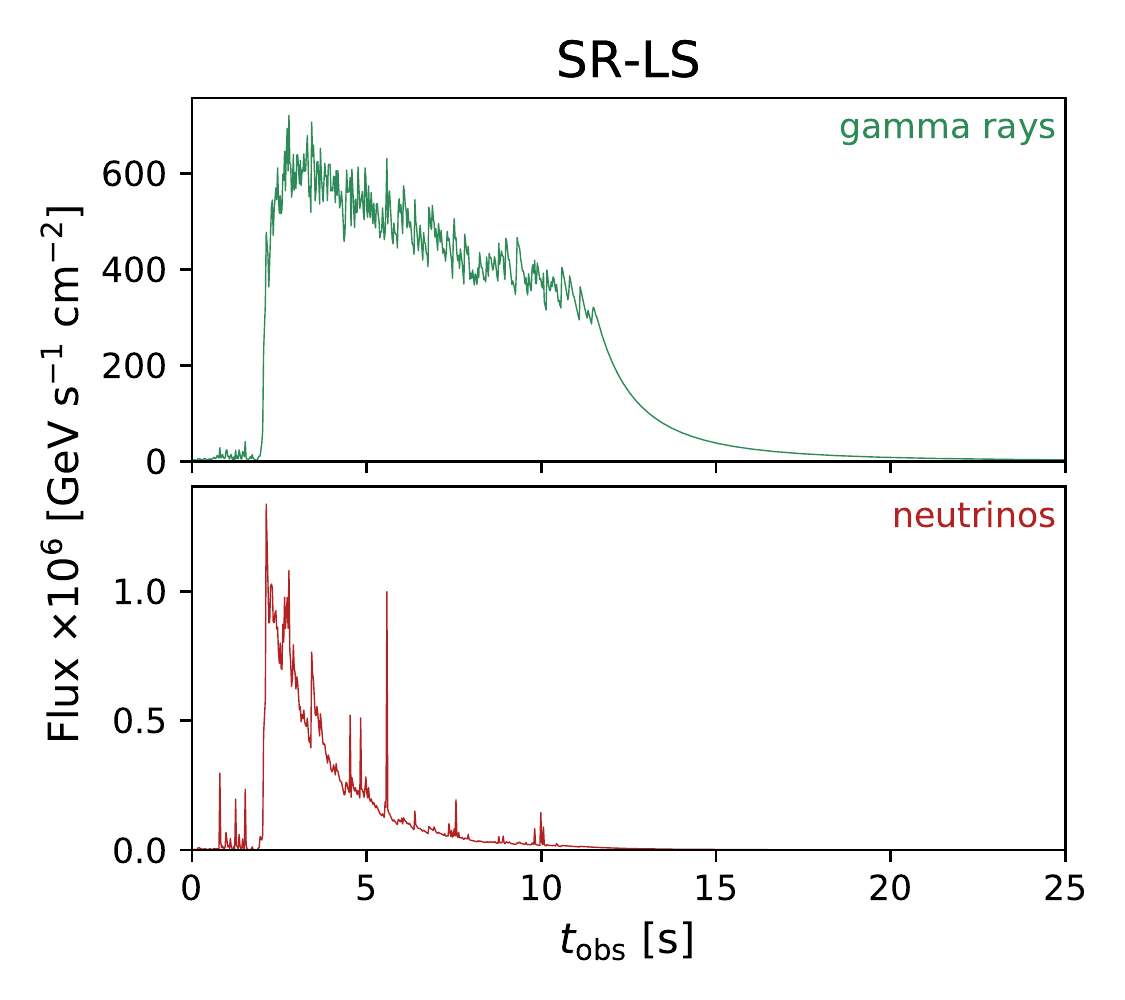} \\
	\includegraphics[width=.35\textwidth]{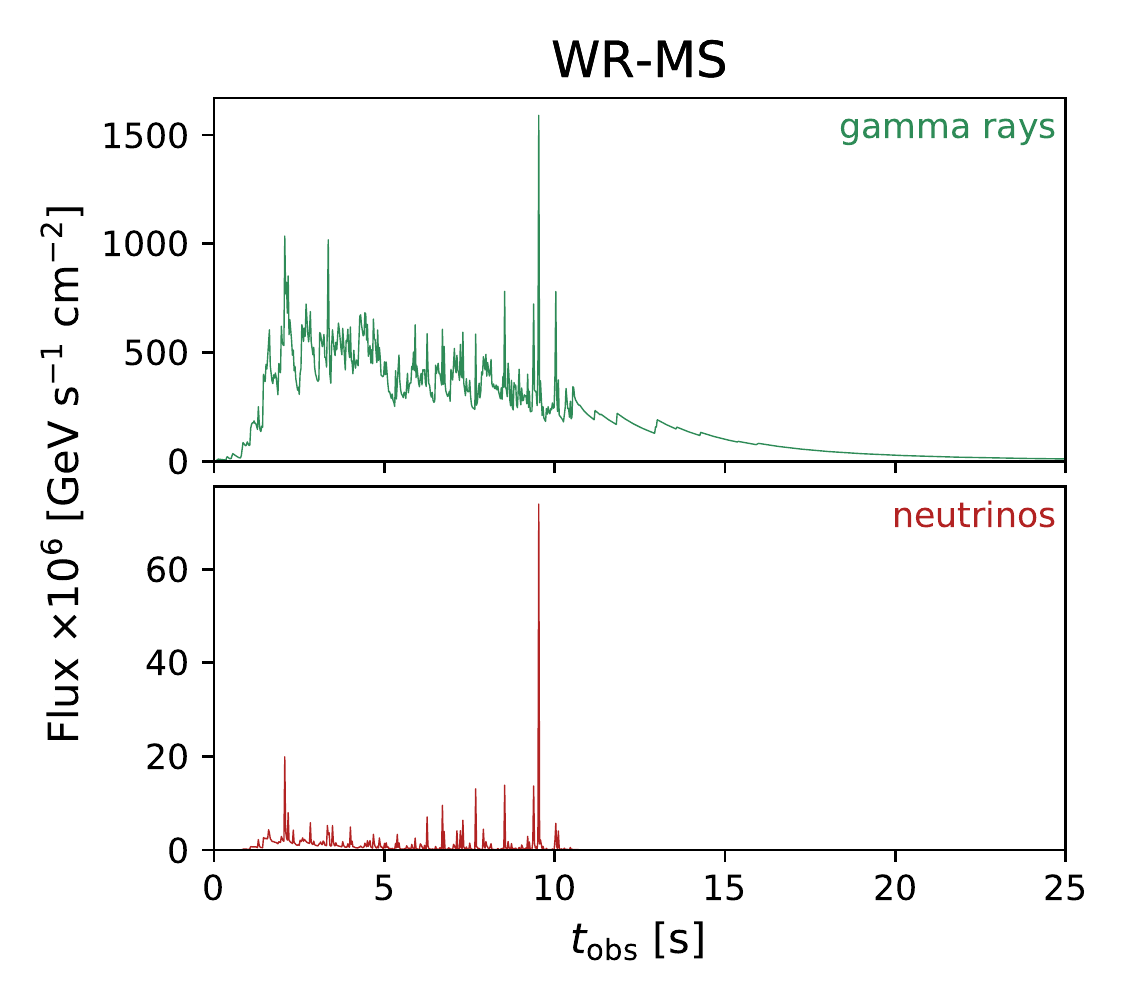} \hspace*{0.05\textwidth}
	\includegraphics[width=.35\textwidth]{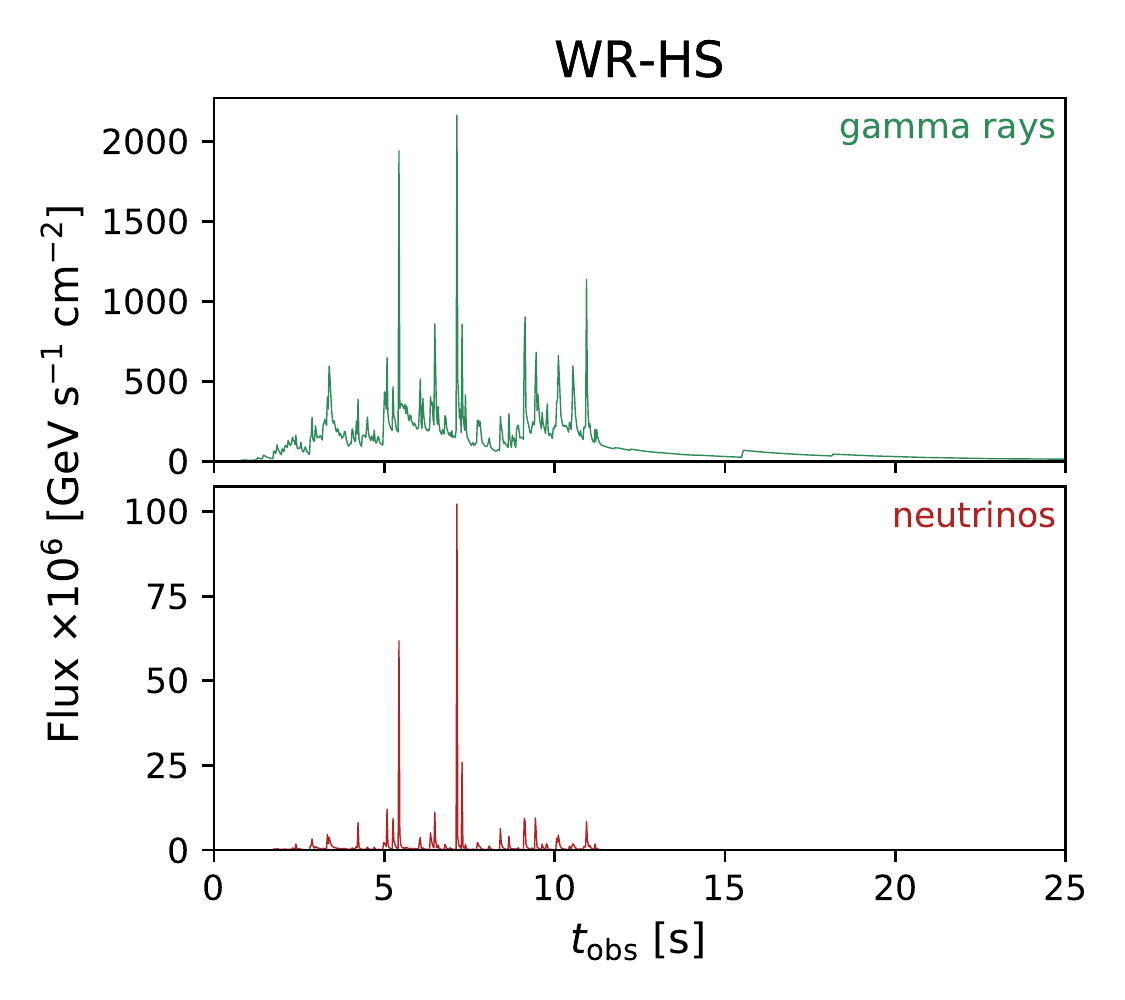}
\caption{\label{fig:light_curves} Resulting light curves (flux as a function of time in gamma ray and neutrinos) for the four example cases defined in \figu{initial_shells} (see figure titles). The light curves are obtained by assuming that each collision emits a fast rise and exponential decay peak (`FRED'), which is normalized to its total energy ouput. Each light curve is shown for a single GRB assuming a redshift of two.}
\end{figure*}

In \figu{light_curves}, we show the light curves, both in gamma rays and neutrinos, for an individual GRB for the four different model assumptions \figu{initial_shells}. Note that these examples are sorted by increasing stochasticity and decreasing engine ramp-up. The case \first{} corresponds to a very disciplined engine in \citet{Bustamante:2016wpu}, leading to a single-pulsed light curve in gamma rays. The increasing engine stochasticity adds time variability to that  light curve, see \second{} and \third{}. In the extreme case \fourth{}, the ramp-up is sub-dominant, the stochasticity of the engine leads to a very spiky light curve, and the underlying (longer) pulse structure is gone. Thus these gamma-ray light curves are clearly qualitatively different, and may be used as model discriminators for the individual GRB. Note that the height of the individual peaks is determined by the luminosity of each collision, which increases with the difference in the Lorentz factors of the colliding shells. 

If the light curves are to be used as model discriminator for the UHECR model, these results need to be compared to the whole population of GRBs. That, however, is not trivial, as GRBs come in a large variety of light curves and there are limitations to resolve the time variability especially for GRBs detected with low statistics; thus it is likely that there are selection effects. For example, while most detected GRBs appear to be single-pulsed with very simple structure, they may be detected at too low statistics to resolve any features or time structure. It is therefore conceivable that the whole GRB population consists of a mix of different light curve types, similar to the ones in \figu{light_curves}. That is a subject beyond the scope of this work which requires further study if light curves are to be used as model-discriminator.

The neutrino light curves are correlated with the gamma-rays, although there is no one-to-one correspondence. The reason for that is that the pion production efficiency scales $\propto R_C^{-2}$ and that the innermost collisions lead to higher neutrino production. In stochastic models (\eg\ \fourth{}) the collision radii are randomly distributed and, in particular, not correlated with observation time. The neutrino peaks are randomly enhanced compared to the gamma ray peaks depending on where that collision occurred.  In deterministic models (\eg\ \first{}), collision radius and observation time are correlated such that the first collisions come from the innermost radii and that the neutrino production efficiency quickly drops with time. In such GRBs, the very-high energetic (TeV) gamma rays can be surpressed early-on because of gamma-gamma pair production, see~\citet{Bustamante:2016wpu} for a more detailed discussion.

One could use these observations for the optimization of the GRB stacking searches: For single-pulsed light curves, the neutrinos are expected to arrive early after the gamma-ray trigger (in the first few seconds for the s), whereas for highly-variable light curves, the neutrinos could arrive at any given time during $T_{90}$.

\subsection{Post-dicted neutrino fluxes from GRBs}

\begin{figure*}
	\includegraphics[width=.47\textwidth]{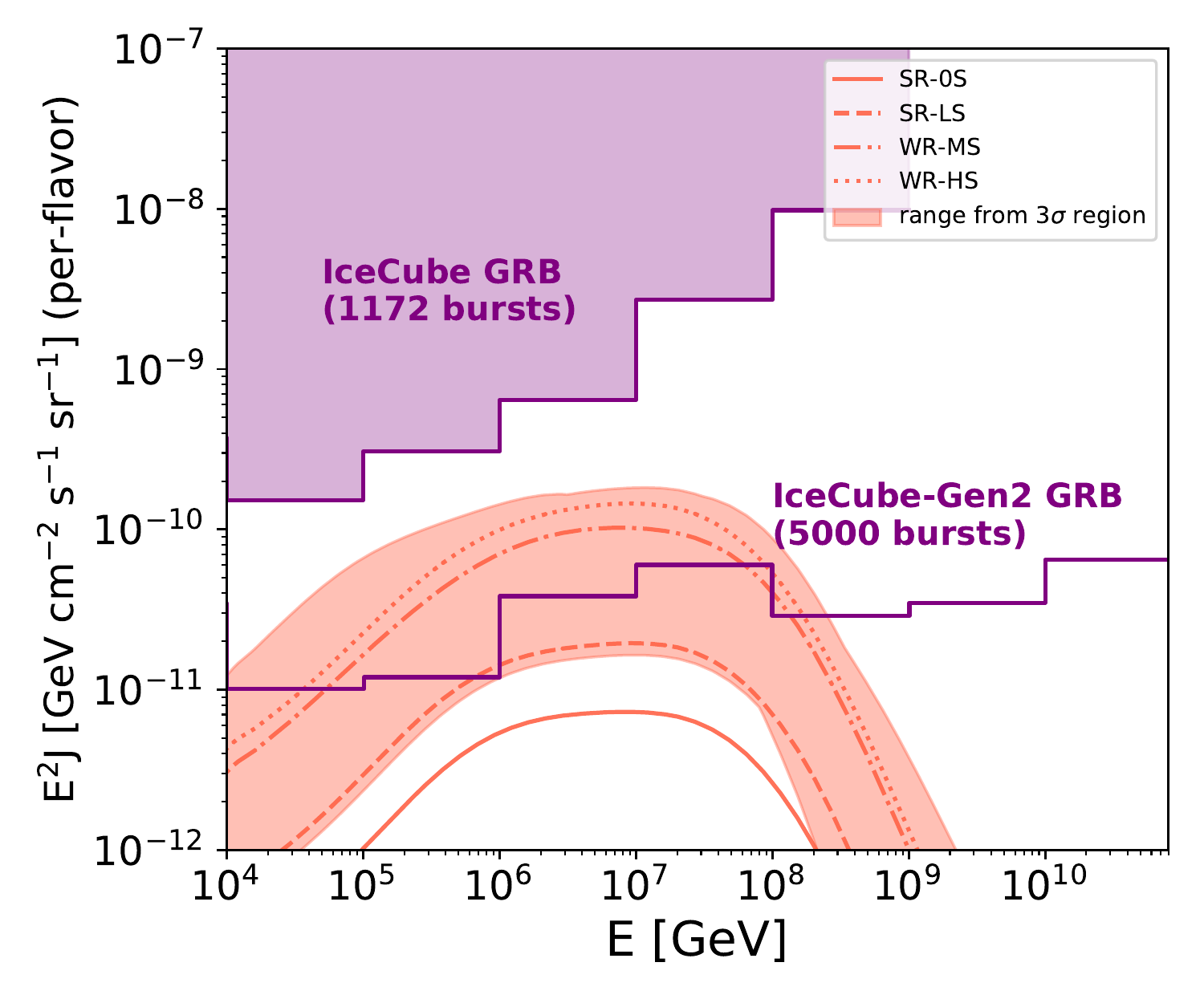}
	\includegraphics[width=.47\textwidth]{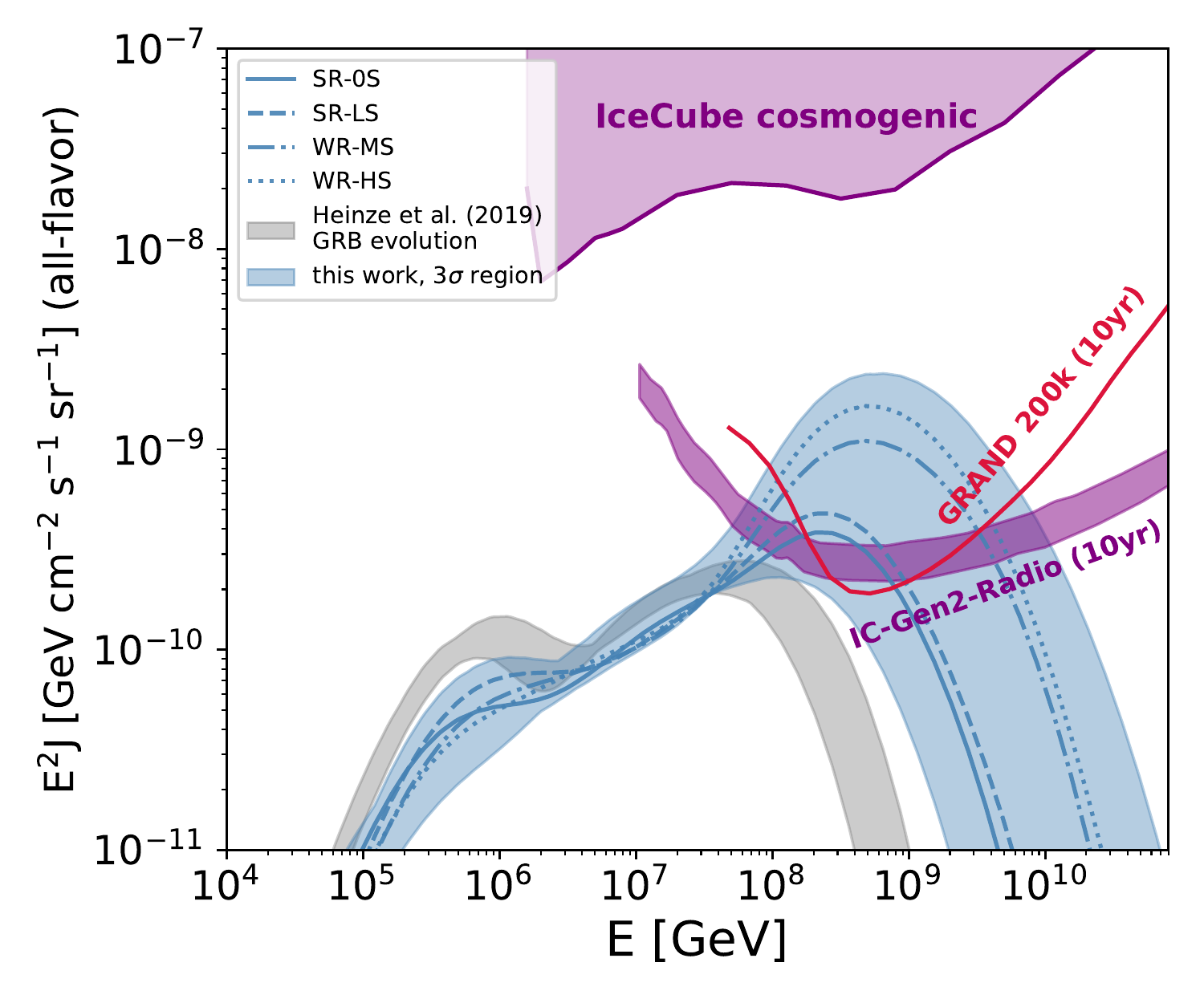}
\caption{\label{fig:neutrino_fluxes}
	Post-dicted single-flavor prompt (left panel) and all-flabor cosmogenic (right panel) neutrino fluxes from the model fit to UHECR data. The shaded regions correspond to the range derived from the UHECR fit ($3 \sigma$-contour in \figu{paramspace}), the different curves to the different setups defined in \figu{initial_shells} as indicated in the plot legend. For comparison, the current IceCube GRB stacking limit for the prompt phase~\citep{Aartsen:2017wea,Aartsen:2020fgd} as well as the projected limit for IceCube-Gen2 (for 5000 bursts)~\citep{Aartsen:2020fgd} are shown in the left panel, and the current cosmogenic neutrino flux limit~\citep{Aartsen:2018vtx} and selected future limits~\citep{Alvarez-Muniz:2018bhp,Aartsen:2020fgd} in the right panel. In the right panel, the expectation for a generic rigidity-dependent UHECR fit is shown for comparison, see \citet[Fig.11]{Heinze:2019jou}.
	}
\end{figure*}

Neutrinos have been proposed in the past as a model discriminator to potentially rule out the UHECR origin from GRBs. We therefore derive the ``post-dicted'' prompt and cosmogenic neutrino fluxes from the $3\sigma$ contour of the UHECR fit in \figu{paramspace}. They are shown for the source/prompt neutrinos and cosmogenic neutrinos as shaded regions in the left and right panels of \figu{neutrino_fluxes}, respectively.

The deterministic engine (\first{}) produces the lowest neutrino flux since the typical collision radius is large, and it is outside of the shaded range, since it is not within the $3 \sigma$ contours. The neutrino flux increases with stochasticity, as there are more collisions at low collision radii (but above the photosphere), see \figu{particle_radii_first} and \figu{particle_radii_third}. 

The post-dicted prompt neutrino fluxes (left panel of \figu{neutrino_fluxes}) are well below the GRB stacking limits from the IceCube Observatory \citep{Aartsen:2017wea}. One way to interpret this result is that even with generous variations among individual GRBs, the absence of neutrino associations in the present detectors does not exclude GRBs as the origin of UHECRs. Our model does not impose exotic fireball parameters, high baryonic loading or excessive Lorentz factors. \revised{The computed neutrino fluxes are derived in a post-dictive way from the UHECR fit without imposing any additional bias from the non-observation of  neutrinos by IceCube, and it is somewhat surprising that no model within the $3 \sigma$ UHECR fit produces a detectable prompt neutrino flux. The fluxes are compatible with earlier estimates in \citet{Bustamante:2014oka,Bustamante:2016wpu,Rudolph:2019ccl} for a fixed baryonic loading, whereas in this work these are obtained from the UHECR fit (see \tabl{examples}) by integrating over the entire source population.} This result demonstrates that the initial interpretation of IceCube's non-observation \citep{Abbasi:2012zw} can be regarded as too strong, and that the UHECR origin from GRBs cannot be ruled out based on neutrino observations, yet.

The next generation detectors, such as  KM3Net-ARCA in the Mediterranean Sea~\citep{Aiello:2018usb}, demonstrate promising full sky sensitivity estimates for point-source detection  and hence may detect some of the GRBs. The planned IceCube-Gen2 detector at the South Pole will also have an enhanced sensitivity to GRBs because of the significantly larger effective area and because the stacking search should be still statistics-limited. It is therefore conceivable that the exposure next-generation of neutrino experiments, in particular that of IceCube-Gen2 as shown in \figu{neutrino_fluxes}, will be sufficient to probe the UHECR paradigm. It is expected that these future detectors can at least exclude the cases with high source stochasticity.

Concerning cosmogenic neutrinos, \citet{Heinze:2019jou} and \citet{AlvesBatista:2018zui} demonstrated that for alike, homogeneously distributed UHECR sources that accelerate nuclei up to a maximal rigidity, the detection of cosmogenic neutrinos from UHECR nuclei is out of reach for the next generation detectors. This statement is valid considering model systematics of the propagation and the air-shower model \citep{Heinze:2019jou}. The present model captures some of the variety of observed light curves by scanning over engine properties. The neutrino fluxes in right panel of \figu{neutrino_fluxes}, therefore, include some non-trivial scenarios with (for example) a high-energy proton component at energies higher than in simple rigidity-dependent sources models. This sub-leading proton contribution increases with the level of stochasticity in our model, see \figu{observed_spectra}. Since these protons reach the threshold for CMB interactions, the cosmogenic neutrino flux is significantly enhanced, see \citet{vanVliet:2019nse} for a detailed discussion. This is prominently visible in the right panel of \figu{neutrino_fluxes} that indicates the possibility to observe a diffuse component from GRB with the next generation radio detectors.

We notice here that the use of the $\sigma(X_\mathrm{max})$ in the fit would reduce those cases in which protons/neutrons reach very high energies in the escape spectra, with the consequence of suppressing the production of neutrinos in the extragalactic space. \revised{In this case, the $3\sigma$ range of the cosmogenic neutrino flux would be barely detectable with ten years of  IceCube-Gen2 or GRAND exposure.} Since the production of prompt neutrinos is mostly dependent on the efficiency of the in-source interactions and not on the maximum energy of the cosmic rays, the result on the source neutrino flux is qualitatively not affected by the use of the $\sigma(X_\mathrm{max})$ in the fit.

\section{Summary and conclusions}

We have performed a systematic parameter space study of the engine properties of GRBs in the internal shock scenario in the context of multi-messenger observations. Since neutrino observations are known to constrain the UHECR origin from GRBs in one-zone models, our model includes multiple internal shocks, and can describe both stochastic engines and deterministic ramp-ups over a wide parameter space. The main target of our study has been the question if long-duration GRBs can describe UHECR spectrum and composition data in spite of existing limits from the non-observation of GRB (prompt and cosmogenic) neutrinos. 

We have demonstrated that UHECR data (spectrum and $\langle X_{\mathrm{max}} \rangle$) can indeed be described in a wide range of engine parameters, and that UHECR data alone cannot discriminate among different options. The observed $\sigma(X_{\mathrm{max}})$, however,  indicates a rather pure composition at the highest energies, which is intrinsically difficult to obtain in multi-region or population models in which a diversity of production regions or sources contribute. Here the superposition of UHECR emissions from different collision radii results in a larget overlap of light and heavy masses than observed in data for the majority of the tested engine parameters. 
We speculate that either $\sigma(X_{\mathrm{max}})$ is more strongly affected by hadronic uncertainties than anticipated, or that significant fine-tuning is needed to describe $\sigma(X_{\mathrm{max}})$ if GRBs are to describe the diffuse UHECR flux -- otherwise GRBs may not be the dominant source of the UHECR flux. A similar argument can be made for other source populations that are dominated by higher redshifts, unless there is a mechanism to fine-tune the maximal rigidity of each source. Therefore the trend in $\sigma(X_{\mathrm{max}})$ might be better described by a small population of local sources.

We have  ``post-dicted'' the neutrino flux from the $3 \sigma$ allowed region of the UHECR fit, and found that the expected neutrino flux is well below the current stacking bound in consistency with earlier estimates from multi-collision models. This means that the UHECR paradigm cannot yet be ruled out with neutrino data alone; we however expect that the next generation of neutrino telescopes could observe GRB neutrinos if GRBs powered the UHECR flux. If no neutrinos are observed, especially stochastic engine models will be effectively constrained with these detectors.

We have also shown that the fraction of nuclei heavier than helium to be injected at the source has to be larger than 70\% (at the 95\% CL) in order to describe the UHECR data. This reflects the distribution of isotopes at the end of the core-collapse, which can be attributed to several different characteristics of the pre-supernova models.

As possible model discriminators, which are sensitive to the stochasticity of the engine, the GRB light curves have been identified. While these can be easily used for individual GRBs, it may be difficult to associate them with a whole population of GRBs.

Furthermore, we have demonstrated that the (isotropic-equivalent) kinetic energy of the outflow has to be $ \sim 10^{55} \, \mathrm{erg} - 10^{56} \, \mathrm{erg}$ per GRB implying a low transfer efficiency into electromagnetic radiation, if the population of long-duration GRBs is powering the UHECRs. The dissipation efficiency problem (transfer from kinetic energy to non-thermal radiation including baryons) is in that case only one part of the problem, the other is how to accelerate baryons to the highest energies without leaving too much energy below the UHECR energy. The energy requirement can be somewhat relaxed for hard (significantly harder than $E^{-2}$) acceleration spectra, or an acceleration mechanism with an extremely high transfer efficiency to the highest energies. Nevertheless, this extreme energy requirement is difficult to circumvent, and may be tested by afterglow observations.

We conclude that the GRB paradigm for UHECRs cannot be uniquely excluded at this point. On the one hand, the description of UHECR data has revealed that spectrum and $\langle X_{\mathrm{max}} \rangle$ can in fact described for a wide range of engine parameters with an inferred bulk Lorentz factor $\simeq 200-400$ within expectations, and the post-dicted prompt and cosmogenic neutrino fluxes are below current limits without any prior or bias. On the other hand, major obstacles have been the description of the observed $\sigma(X_{\mathrm{max}})$ and the required energetics. 
Only independent observations, such as limits or detections from the next generation of neutrino detectors, will be able to robustly exclude the GRB-UHECR connection.

\subsection*{Acknowledgments}

We thank No\'emie Globus, Arjen van Vliet and Markus Ackermann for useful discussions. DB acknowledges the participation to the Pierre Auger Collaboration. 

This work has been supported by the European Research Council (ERC) under the European Union's Horizon 2020 research and innovation programme (Grant No. 646623). 
The work was supported by the International Helmholtz-Weizmann Research School for Multimessenger Astronomy, largely funded through the Initiative and Networking Fund of the Helmholtz Association.
AF completed parts of this work as JSPS International Research Fellow (JSPS KAKENHI Grant Number 19F19750).

\subsection*{Data availability}
There are no new data associated with this article.



\bibliographystyle{mnras}
\bibliography{references} 

\begin{thebibliography}{}
\makeatletter
\relax
\def\mn@urlcharsother{\let\do\@makeother \do\$\do\&\do\#\do\^\do\_\do\%\do\~}
\def\mn@doi{\begingroup\mn@urlcharsother \@ifnextchar [ {\mn@doi@}
  {\mn@doi@[]}}
\def\mn@doi@[#1]#2{\def\@tempa{#1}\ifx\@tempa\@empty \href
  {http://dx.doi.org/#2} {doi:#2}\else \href {http://dx.doi.org/#2} {#1}\fi
  \endgroup}
\def\mn@eprint#1#2{\mn@eprint@#1:#2::\@nil}
\def\mn@eprint@arXiv#1{\href {http://arxiv.org/abs/#1} {{\tt arXiv:#1}}}
\def\mn@eprint@dblp#1{\href {http://dblp.uni-trier.de/rec/bibtex/#1.xml}
  {dblp:#1}}
\def\mn@eprint@#1:#2:#3:#4\@nil{\def\@tempa {#1}\def\@tempb {#2}\def\@tempc
  {#3}\ifx \@tempc \@empty \let \@tempc \@tempb \let \@tempb \@tempa \fi \ifx
  \@tempb \@empty \def\@tempb {arXiv}\fi \@ifundefined
  {mn@eprint@\@tempb}{\@tempb:\@tempc}{\expandafter \expandafter \csname
  mn@eprint@\@tempb\endcsname \expandafter{\@tempc}}}

\bibitem[\protect\citeauthoryear{Aab et~al.}{Aab et~al.}{2017}]{Aab:2016zth}
Aab A.,  et~al., 2017, \mn@doi [JCAP] {10.1088/1475-7516/2017/04/038}, 1704,
  038

\bibitem[\protect\citeauthoryear{Aab et~al.}{Aab et~al.}{2018}]{Aab:2018chp}
Aab A.,  et~al., 2018, \mn@doi [Astrophys. J.] {10.3847/2041-8213/aaa66d}, 853,
  L29

\bibitem[\protect\citeauthoryear{Aab et~al.}{Aab et~al.}{2019}]{Aab:2019auo}
Aab A.,  et~al., 2019, \mn@doi [JCAP] {10.1088/1475-7516/2019/10/022}, 1910,
  022

\bibitem[\protect\citeauthoryear{Aartsen et~al.}{Aartsen
  et~al.}{2017}]{Aartsen:2017wea}
Aartsen M.~G.,  et~al., 2017, \mn@doi [Astrophys. J.]
  {10.3847/1538-4357/aa7569}, 843, 112

\bibitem[\protect\citeauthoryear{Aartsen et~al.}{Aartsen
  et~al.}{2018}]{Aartsen:2018vtx}
Aartsen M.~G.,  et~al., 2018, \mn@doi [Phys. Rev.]
  {10.1103/PhysRevD.98.062003}, D98, 062003

\bibitem[\protect\citeauthoryear{Aartsen et~al.}{Aartsen
  et~al.}{2020}]{Aartsen:2020fgd}
Aartsen M.,  et~al., 2020, arXiv:2008.04323

\bibitem[\protect\citeauthoryear{Abbasi et~al.}{Abbasi
  et~al.}{2012}]{Abbasi:2012zw}
Abbasi R.,  et~al., 2012, \mn@doi [Nature] {10.1038/nature11068}, 484, 351

\bibitem[\protect\citeauthoryear{Abdalla et~al.}{Abdalla
  et~al.}{2019}]{Arakawa:2019cfc}
Abdalla H.,  et~al., 2019, \mn@doi [Nature] {10.1038/s41586-019-1743-9}, 575,
  464

\bibitem[\protect\citeauthoryear{Abreu et~al.}{Abreu
  et~al.}{2013}]{Abreu:2013env}
Abreu P.,  et~al., 2013, \mn@doi [JCAP] {10.1088/1475-7516/2013/02/026}, 1302,
  026

\bibitem[\protect\citeauthoryear{Acciari et~al.}{Acciari
  et~al.}{2019}]{Acciari:2019dxz}
Acciari V.~A.,  et~al., 2019, \mn@doi [Nature] {10.1038/s41586-019-1750-x},
  575, 455

\bibitem[\protect\citeauthoryear{Aiello et~al.}{Aiello
  et~al.}{2019}]{Aiello:2018usb}
Aiello S.,  et~al., 2019, \mn@doi [Astropart. Phys.]
  {10.1016/j.astropartphys.2019.04.002}, 111, 100

\bibitem[\protect\citeauthoryear{Alvarez-Muniz et~al.}{Alvarez-Muniz
  et~al.}{2020}]{Alvarez-Muniz:2018bhp}
Alvarez-Muniz J.,  et~al., 2020, \mn@doi [Sci. China Phys. Mech. Astron.]
  {10.1007/s11433-018-9385-7}, 63, 219501

\bibitem[\protect\citeauthoryear{Alves~Batista, Boncioli, di Matteo, van Vliet
  \& Walz}{Alves~Batista et~al.}{2015}]{Batista:2015mea}
Alves~Batista R.,  Boncioli D.,  di Matteo A.,  van Vliet A.,   Walz D.,  2015,
  \mn@doi [JCAP] {10.1088/1475-7516/2015/10/063}, 1510, 063

\bibitem[\protect\citeauthoryear{Alves~Batista, de Almeida, Lago  \&
  Kotera}{Alves~Batista et~al.}{2019}]{AlvesBatista:2018zui}
Alves~Batista R.,  de Almeida R.~M.,  Lago B.,   Kotera K.,  2019, \mn@doi
  [JCAP] {10.1088/1475-7516/2019/01/002}, 1901, 002

\bibitem[\protect\citeauthoryear{Baerwald, Bustamante  \& Winter}{Baerwald
  et~al.}{2013}]{Baerwald:2013pu}
Baerwald P.,  Bustamante M.,   Winter W.,  2013, \mn@doi [Astrophys. J.]
  {10.1088/0004-637X/768/2/186}, 768, 186

\bibitem[\protect\citeauthoryear{Baerwald, Bustamante  \& Winter}{Baerwald
  et~al.}{2015}]{Baerwald:2014zga}
Baerwald P.,  Bustamante M.,   Winter W.,  2015, Astropart. Phys., 62, 66

\bibitem[\protect\citeauthoryear{{Bellido, J. et al.}}{{Bellido, J. et
  al.}}{2017}]{Bellido:2017}
{Bellido, J. et al.} 2017. PoS(ICRC2017)506

\bibitem[\protect\citeauthoryear{Beloborodov}{Beloborodov}{2017}]{Beloborodov:2016jmz}
Beloborodov A.~M.,  2017, \mn@doi [Astrophys. J.] {10.3847/1538-4357/aa5c8c},
  838, 125

\bibitem[\protect\citeauthoryear{Beniamini, Nava, Barniol~Duran  \&
  Piran}{Beniamini et~al.}{2015}]{Beniamini:2015eaa}
Beniamini P.,  Nava L.,  Barniol~Duran R.,   Piran T.,  2015, \mn@doi [Mon.
  Not. Roy. Astron. Soc.] {10.1093/mnras/stv2033}, 454, 1073

\bibitem[\protect\citeauthoryear{Beniamini, Nava  \& Piran}{Beniamini
  et~al.}{2016}]{Beniamini:2016hzc}
Beniamini P.,  Nava L.,   Piran T.,  2016, \mn@doi [Mon. Not. Roy. Astron.
  Soc.] {10.1093/mnras/stw1331}, 461, 51

\bibitem[\protect\citeauthoryear{Berezinsky \& Zatsepin}{Berezinsky \&
  Zatsepin}{1969}]{Beresinsky:1969qj}
Berezinsky V.~S.,  Zatsepin G.~T.,  1969, \mn@doi [Phys. Lett.]
  {10.1016/0370-2693(69)90341-4}, 28B, 423

\bibitem[\protect\citeauthoryear{Biehl, Boncioli, Fedynitch  \& Winter}{Biehl
  et~al.}{2018}]{Biehl:2017zlw}
Biehl D.,  Boncioli D.,  Fedynitch A.,   Winter W.,  2018, \mn@doi [Astron.
  Astrophys.] {10.1051/0004-6361/201731337}, 611, A101

\bibitem[\protect\citeauthoryear{Boncioli, Fedynitch  \& Winter}{Boncioli
  et~al.}{2017}]{Boncioli:2016lkt}
Boncioli D.,  Fedynitch A.,   Winter W.,  2017, Scientific Reports, 7, 4882

\bibitem[\protect\citeauthoryear{Boncioli, Biehl  \& Winter}{Boncioli
  et~al.}{2019}]{Boncioli:2018lrv}
Boncioli D.,  Biehl D.,   Winter W.,  2019, \mn@doi [Astrophys. J.]
  {10.3847/1538-4357/aafda7}, 872, 110

\bibitem[\protect\citeauthoryear{Bosnjak, Daigne  \& Dubus}{Bosnjak
  et~al.}{2009}]{Bosnjak:2008bd}
Bosnjak Z.,  Daigne F.,   Dubus G.,  2009, \mn@doi [Astron. Astrophys.]
  {10.1051/0004-6361/200811375}, 498, 677

\bibitem[\protect\citeauthoryear{Bo\v{s}njak \& Daigne}{Bo\v{s}njak \&
  Daigne}{2014}]{Bosnjak:2014hya}
Bo\v{s}njak v.,  Daigne F.,  2014, \mn@doi [Astron. Astrophys.]
  {10.1051/0004-6361/201322341}, 568, A45

\bibitem[\protect\citeauthoryear{Bustamante, Baerwald, Murase  \&
  Winter}{Bustamante et~al.}{2015}]{Bustamante:2014oka}
Bustamante M.,  Baerwald P.,  Murase K.,   Winter W.,  2015, \mn@doi [Nature
  Commun.] {10.1038/ncomms7783}, 6, 6783

\bibitem[\protect\citeauthoryear{Bustamante, Murase, Winter  \&
  Heinze}{Bustamante et~al.}{2017}]{Bustamante:2016wpu}
Bustamante M.,  Murase K.,  Winter W.,   Heinze J.,  2017, \mn@doi [Astrophys.
  J.] {10.3847/1538-4357/837/1/33}, 837, 33

\bibitem[\protect\citeauthoryear{Daigne \& Mochkovitch}{Daigne \&
  Mochkovitch}{1998}]{Daigne:1998xc}
Daigne F.,  Mochkovitch R.,  1998, \mn@doi [Mon. Not. Roy. Astron. Soc.]
  {10.1046/j.1365-8711.1998.01305.x}, 296, 275

\bibitem[\protect\citeauthoryear{Daigne, Bosnjak  \& Dubus}{Daigne
  et~al.}{2011}]{Daigne:2010fb}
Daigne F.,  Bosnjak Z.,   Dubus G.,  2011, \mn@doi [Astron. Astrophys.]
  {10.1051/0004-6361/201015457}, 526, A110

\bibitem[\protect\citeauthoryear{Fan \& Piran}{Fan \& Piran}{2006}]{Fan:2006sx}
Fan Y.-H.,  Piran T.,  2006, \mn@doi [Mon. Not. Roy. Astron. Soc.]
  {10.1111/j.1365-2966.2006.10280.x}, 369, 197

\bibitem[\protect\citeauthoryear{{Fenu, F. et al.}}{{Fenu, F. et
  al.}}{2017}]{Fenu:2017}
{Fenu, F. et al.} 2017. PoS(ICRC2017)486

\bibitem[\protect\citeauthoryear{Ghirlanda et~al.,}{Ghirlanda
  et~al.}{2018}]{Ghirlanda:2017opl}
Ghirlanda G.,  et~al., 2018, \mn@doi [Astron. Astrophys.]
  {10.1051/0004-6361/201731598}, 609, A112

\bibitem[\protect\citeauthoryear{Gilmore, Somerville, Primack  \&
  Dominguez}{Gilmore et~al.}{2012}]{Gilmore:2011ks}
Gilmore R.~C.,  Somerville R.~S.,  Primack J.~R.,   Dominguez A.,  2012,
  \mn@doi [Mon. Not. Roy. Astron. Soc.] {10.1111/j.1365-2966.2012.20841.x},
  422, 3189

\bibitem[\protect\citeauthoryear{Globus, Allard, Mochkovitch  \&
  Parizot}{Globus et~al.}{2015a}]{Globus:2014fka}
Globus N.,  Allard D.,  Mochkovitch R.,   Parizot E.,  2015a, \mn@doi [Mon.
  Not. Roy. Astron. Soc.] {10.1093/mnras/stv893}, 451, 751

\bibitem[\protect\citeauthoryear{Globus, Allard  \& Parizot}{Globus
  et~al.}{2015b}]{Globus:2015xga}
Globus N.,  Allard D.,   Parizot E.,  2015b, \mn@doi [Phys. Rev.]
  {10.1103/PhysRevD.92.021302}, D92, 021302

\bibitem[\protect\citeauthoryear{Gottlieb, Levinson  \& Nakar}{Gottlieb
  et~al.}{2020}]{Gottlieb:2020ifs}
Gottlieb O.,  Levinson A.,   Nakar E.,  2020, \mn@doi [Mon. Not. Roy. Astron.
  Soc.] {10.1093/mnras/staa1216}, 495, 570

\bibitem[\protect\citeauthoryear{Gruber et~al.}{Gruber
  et~al.}{2014}]{Gruber:2014iza}
Gruber D.,  et~al., 2014, \mn@doi [Astrophys. J. Suppl.]
  {10.1088/0067-0049/211/1/12}, 211, 12

\bibitem[\protect\citeauthoryear{Heinze, Fedynitch, Boncioli  \& Winter}{Heinze
  et~al.}{2019}]{Heinze:2019jou}
Heinze J.,  Fedynitch A.,  Boncioli D.,   Winter W.,  2019, \mn@doi [Astrophys.
  J.] {10.3847/1538-4357/ab05ce}, 873, 88

\bibitem[\protect\citeauthoryear{H{\"u}mmer, Baerwald  \& Winter}{H{\"u}mmer
  et~al.}{2012}]{Hummer:2011ms}
H{\"u}mmer S.,  Baerwald P.,   Winter W.,  2012, \mn@doi [Phys. Rev. Lett.]
  {10.1103/PhysRevLett.108.231101}, 108, 231101

\bibitem[\protect\citeauthoryear{{James} \& {Roos}}{{James} \&
  {Roos}}{1975}]{1975CoPhC..10..343J}
{James} F.,  {Roos} M.,  1975, \mn@doi [Computer Physics Communications]
  {10.1016/0010-4655(75)90039-9}, \href
  {http://adsabs.harvard.edu/abs/1975CoPhC..10..343J} {10, 343}

\bibitem[\protect\citeauthoryear{Kobayashi, Piran  \& Sari}{Kobayashi
  et~al.}{1997}]{Kobayashi:1997jk}
Kobayashi S.,  Piran T.,   Sari R.,  1997, \mn@doi [Astrophys. J.]
  {10.1086/512791}, 490, 92

\bibitem[\protect\citeauthoryear{Koning, Hilaire  \& Duijvestijn}{Koning
  et~al.}{2007}]{Koning:2007}
Koning A.~J.,  Hilaire S.,   Duijvestijn M.~C.,  2007, in {Proceedings,
  International Conference on Nuclear Data for Science and Tecnology}. pp
  211--214

\bibitem[\protect\citeauthoryear{Liang, Zhang  \& Dai}{Liang
  et~al.}{2007}]{Liang:2006ci}
Liang E.,  Zhang B.,   Dai Z.~G.,  2007, \mn@doi [Astrophys. J.]
  {10.1086/517959}, 662, 1111

\bibitem[\protect\citeauthoryear{Liang, Yi, Zhang, LV, Zhang  et~al.}{Liang
  et~al.}{2010}]{Liang:2009zi}
Liang E.-W.,  Yi S.-x.,  Zhang J.,  LV H.-J.,  Zhang B.-B.,   et~al., 2010,
  \mn@doi [Astrophys. J.] {10.1088/0004-637X/725/2/2209}, 725, 2209

\bibitem[\protect\citeauthoryear{Murase \& Nagataki}{Murase \&
  Nagataki}{2006}]{Murase:2005hy}
Murase K.,  Nagataki S.,  2006, \mn@doi [Phys. Rev. D]
  {10.1103/PhysRevD.73.063002}, 73, 063002

\bibitem[\protect\citeauthoryear{Murase, Ioka, Nagataki  \& Nakamura}{Murase
  et~al.}{2008}]{Murase:2008mr}
Murase K.,  Ioka K.,  Nagataki S.,   Nakamura T.,  2008, \mn@doi [Phys.Rev.]
  {10.1103/PhysRevD.78.023005}, D78, 023005

\bibitem[\protect\citeauthoryear{Ohira, Murase  \& Yamazaki}{Ohira
  et~al.}{2010}]{Ohira:2009rd}
Ohira Y.,  Murase K.,   Yamazaki R.,  2010, \mn@doi [Astron. Astrophys.]
  {10.1051/0004-6361/200913495}, 513, A17

\bibitem[\protect\citeauthoryear{{Pohl}}{{Pohl}}{1993}]{1993A&A...270...91P}
{Pohl} M.,  1993, Astron. Astrophys., \href
  {https://ui.adsabs.harvard.edu/abs/1993A&A...270...91P} {270, 91}

\bibitem[\protect\citeauthoryear{Rees \& Meszaros}{Rees \&
  Meszaros}{1992}]{Rees:1992ek}
Rees M.,  Meszaros P.,  1992, Mon. Not. Roy. Astron. Soc., 258, 41

\bibitem[\protect\citeauthoryear{Rees \& Meszaros}{Rees \&
  Meszaros}{1994}]{Rees:1994nw}
Rees M.,  Meszaros P.,  1994, Astrophys. J., 430, L93

\bibitem[\protect\citeauthoryear{Rudolph, Heinze, Fedynitch  \& Winter}{Rudolph
  et~al.}{2020}]{Rudolph:2019ccl}
Rudolph A.,  Heinze J.,  Fedynitch A.,   Winter W.,  2020, \mn@doi [Astrophys.
  J.] {10.3847/1538-4357/ab7ea7}, 893, 72

\bibitem[\protect\citeauthoryear{Samuelsson, Bégué, Ryde  \&
  Pe'er}{Samuelsson et~al.}{2019}]{Samuelsson:2018fan}
Samuelsson F.,  Bégué D.,  Ryde F.,   Pe'er A.,  2019, \mn@doi [Astrophys.
  J.] {10.3847/1538-4357/ab153c}, 876, 93

\bibitem[\protect\citeauthoryear{Sironi \& Spitkovsky}{Sironi \&
  Spitkovsky}{2011}]{Sironi:2010rb}
Sironi L.,  Spitkovsky A.,  2011, \mn@doi [Astrophys.J.]
  {10.1088/0004-637X/726/2/75}, 726, 75

\bibitem[\protect\citeauthoryear{Vietri}{Vietri}{1995}]{vietri1995acceleration}
Vietri M.,  1995, arXiv preprint astro-ph/9506081

\bibitem[\protect\citeauthoryear{Wanderman \& Piran}{Wanderman \&
  Piran}{2010}]{Wanderman:2009es}
Wanderman D.,  Piran T.,  2010, Mon.Not.Roy.Astron.Soc., 406, 1944

\bibitem[\protect\citeauthoryear{Waxman}{Waxman}{1995}]{waxman1995cosmological}
Waxman E.,  1995, Physical Review Letters, 75, 386

\bibitem[\protect\citeauthoryear{Waxman \& Bahcall}{Waxman \&
  Bahcall}{1997}]{Waxman:1997ti}
Waxman E.,  Bahcall J.~N.,  1997, \mn@doi [Phys. Rev. Lett.]
  {10.1103/PhysRevLett.78.2292}, 78, 2292

\bibitem[\protect\citeauthoryear{Woosley \& Heger}{Woosley \&
  Heger}{2006}]{Woosley:2005gy}
Woosley S.,  Heger A.,  2006, \mn@doi [Astrophys.\ J.] {10.1086/498500}, 637,
  914

\bibitem[\protect\citeauthoryear{Zhang, Murase, Kimura, Horiuchi  \&
  Meszaros}{Zhang et~al.}{2018}]{Zhang:2017moz}
Zhang B.~T.,  Murase K.,  Kimura S.~S.,  Horiuchi S.,   Meszaros P.,  2018,
  \mn@doi [Phys. Rev.] {10.1103/PhysRevD.97.083010}, D97, 083010

\bibitem[\protect\citeauthoryear{van Vliet, Alves~Batista  \& Horandel}{van
  Vliet et~al.}{2019}]{vanVliet:2019nse}
van Vliet A.,  Alves~Batista R.,   Horandel J.~R.,  2019, \mn@doi [Phys. Rev.]
  {10.1103/PhysRevD.100.021302}, D100, 021302

\makeatother
\end{thebibliography}

\bsp	
\label{lastpage}
\end{document}